\documentclass[a4paper,11pt]{article}
\usepackage{jcappub} % for details on the use of the package, please see the JINST-author-manual
\usepackage{lineno}
\usepackage{graphicx}% Include figure files
\usepackage{dcolumn}% Align table columns on decimal point
\usepackage{bm}% bold math
\usepackage{amsmath,amssymb} 
\usepackage{amsfonts}
\usepackage{subfigure}
\usepackage{graphicx}
\usepackage{ulem}
\usepackage{soul}
\usepackage{color}
\usepackage{xcolor}
\usepackage{float}
\usepackage{makecell}
\usepackage{newtxtext,newtxmath}
\usepackage{mathrsfs}
\usepackage{gensymb}
\usepackage{multirow}
\usepackage{booktabs}
\usepackage{threeparttable}
\usepackage{ulem}
\usepackage{rotating}
\usepackage{graphicx}
\usepackage{tabularx}
\usepackage{verbatim}
\usepackage{url}
\usepackage{hyperref}% add hypertext capabilities

\arxivnumber{2411.14510} % Only if you have one
\title{AKRA 2.0: Accurate Kappa Reconstruction Algorithm for masked shear catalog }

\newcommand{\m}{\ensuremath{\, {\rm {m}}}}

\author[a,b]{Yuan Shi}
\emailAdd{yshi@sjtu.edu.cn}

\author[a,b,c]{Pengjie Zhang}
\emailAdd{zhangpj@sjtu.edu.cn}

\author[d,e,f]{Furen Deng}
\author[a,b]{Shuren Zhou}
\author[a,b]{Hongbo Cai}
\author[g]{Ji Yao}
\author[h]{Zeyang Sun}

\affiliation[a]{Department of Astronomy, School of Physics and Astronomy, Shanghai Jiao Tong University, Shanghai, 200240, China}
\affiliation[b]{Key Laboratory for Particle Astrophysics and Cosmology (MOE) / Shanghai Key Laboratory for Particle Physics and Cosmology, China}
\affiliation[c]{Division of Astronomy and Astrophysics, Tsung-Dao Lee Institute, Shanghai Jiao Tong University, Shanghai, 200240, China}
\affiliation[d]{National Astronomical Observatories, Chinese Academy of Sciences, 20A Datun Road, Beijing 100101, P. R. China}
\affiliation[e]{University of Chinese Academy of Sciences, School of Astronomy and Space Science, Beijing 100049, P. R. China}
\affiliation[f]{Institute of Astronomy, University of Cambridge, Madingley Road, Cambridge, CB3 0HA, UK}
\affiliation[g]{Shanghai Astronomical Observatory (SHAO), Nandan Road 80, Shanghai, China}
\affiliation[h]{Universitäts-Sternwarte, Fakultät für Physik, Ludwig-Maximilians Universität München, Scheinerstr. 1, 81679 München, Germany}

\abstract{Cosmic shear surveys serve as a powerful tool for mapping the underlying matter density field, including non-visible dark matter. 
    A key challenge in cosmic shear surveys is the accurate reconstruction of lensing convergence ($\kappa$) maps from shear catalogs impacted by survey boundaries and masks, which seminal Kaiser-Squires (KS) method are not designed to handle. To overcome these limitations, we previously proposed the Accurate Kappa Reconstruction
    Algorithm (AKRA), a prior-free maximum likelihood map-making method.
    Initially designed for flat sky scenarios with periodic boundary conditions, AKRA has proven successful in recovering high-precision $\kappa$ maps from masked shear catalogs.
    In this work, we upgrade AKRA to AKRA 2.0 by integrating the tools designed for spherical geometry. This upgrade employs spin-weighted spherical harmonic transforms to reconstruct the convergence field over the full sky.  To optimize computational efficiency, we implement a scale-splitting strategy that segregates the analysis into two parts: large-scale analysis on the sphere (referred to as AKRA-sphere) and small-scale analysis on the flat sky (referred to as AKRA-flat); the results from both analyses are then combined to produce  final reconstructed $\kappa$ map.
    We tested AKRA 2.0 using simulated shear catalogs with various masks, demonstrating that the reconstructed $\kappa$ map by AKRA 2.0 maintains high accuracy. For the reconstructed $\kappa$ map in unmasked regions, the reconstructed convergence power spectrum $C_\kappa^{\rm{rec}}$ and the correlation coefficient with the true $\kappa$ map $r_\ell$ achieve accuracies of $(1-C_\ell^{\rm{rec}}/C_\ell^{\rm{true}}) \lesssim 1\%$ and $(1-r_\ell) \lesssim 1\%$, respectively.
    Our algorithm is capable of straightforwardly handling further issues such as inhomogeneous shape measurement noise, which we will address in subsequent analysis.}
\begin{document}
\maketitle
\flushbottom

\section{Introduction} \label{sec:intro}
Weak gravitational lensing (WL) is a powerful tool in understanding fundamental cosmological physics, including the nature of dark matter, dark energy, gravity, and astrophysics such as the halo abundance/profile and galaxy-halo connection \cite{Bartelmann2001, Refregier2003, Munshi2008, Kilbinger2015}. It provides a direct probe of the distribution of dark matter, and is pivotal in studying the large-scale structure of the universe in modern galaxy surveys. 
Recent advancements in gravitational lensing have achieved high signal-to-noise (S/N) ratios in cosmic shear measurements from stage III surveys, including the Dark Energy Survey (DES \cite{Abbott2016}), the Hyper Suprime-Cam Survey (HSC \cite{Aihara2018}), and the Kilo-Degree Survey (KiDS \cite{Kuijken2015}). 
Future surveys like the China Space Station Telescope (CSST \cite{Gong2019,Yao2023CSST}), Euclid (\cite{Laureijs2011}), the Rubin Observatory (previously known as the Legacy Survey of Space and Time, LSST \cite{LSST2009}), and the Roman Space Telescope (formerly WFIRST \cite{Spergel2015}), are expected to increase S/N ratio by more than an order of magnitude, reaching S/N over 400 \cite{Yao2023CSST}.

Cosmic shear analyses typically use the two-point correlation function \cite{Martin2013, Troxel2018, Hikage2019, Hamana2020, Asgari2021, Secco2022, LiXiangchong2023} or the power spectrum \cite{Hikage2011, Heymans2012, Lin2012, Kohlinger2017, Hikage2019, Nicola2021, Camacho2021} to extract cosmological constraints. As S/N improves, there is a growing interest in probing the high order statistics, in order to utilize the cosmological information from the non-Gaussianity driven by the non-linear structure formation.
These statistics include peak/void counting statistics \cite{Fan2007, Fan2010, Hilbert2012, Shan2014, Shan2018, Lin2015, Martinet2018, Peel2018, Ajani2020, Martinet2021B, Martinet2021A, Emma2022, Liu2023}, higher-order moments \cite{VanWaerbeke2013, Petri2015, Vicinanza2016, Vicinanza2018, Chang2018, Peel2018}, Minkowski functionals \cite{Kratochvil2012, Petri2015, Vicinanza2019, Parroni2020, Zurcher2021} machine-learning methods \cite{Gupta2018,Ribli2019a,Ribli2019b,Fluri2019,Fluri2022,Matilla2020,LuTianhuan2023} and the direct field-level inferences \cite{PhysRevD.110.023539}.

Higher-order statistics in weak lensing analyses often require a robust algorithm to reconstruct the convergence map from cosmic shear catalogs, also known as mass mapping. 
A seminal and most widely-used tool is the Kaiser-Squires (KS) algorithm \cite{Kaiser1993, Waerbeke2013, Vikram2015, Gatti2022}, which reconstructs the convergence field by assuming scalar convergence with null B-mode. However, it fails to address the impacts introduced by survey boundaries and masks. 
To overcome these limitations, several approaches have been developed.
These include forward modeling of the shear field with a Bayesian prior on the convergence field \cite{Alsing2016, Alsing2017, Porqueres2022}, Wiener filtering \cite{Jeffrey2018}, the use of sparsity and log-normal priors \cite{Leonard2014, Price2019, Jeffrey2018, Fiedorowicz2022b, Fiedorowicz2022a}, and wavelet-based techniques \cite{Starck2006, Starck2021}. 
Moreover, machine learning-based approaches have emerged as promising tools for reconstructing the convergence field, such as the Generative Neural Network \cite{Shirasaki2021} and U-Net-based methods \cite{Jeffrey2020a}. 

Accurate mass mapping remains challenging due to survey boundaries and masks. To solve this rigorously, we proposed the Accurate Kappa Reconstruction Algorithm (AKRA) \cite{shiAKRA2024a}, an unbiased minimum variance linear estimator of the weak lensing convergence field on the flat sky. It has been proven successful in reconstructing high-precision convergence maps on the flat sky. Nevertheless, for a complete solution, analysis for full sky is necessary.

In this work, we introduce AKRA 2.0, an upgraded version of our original AKRA-flat algrithm for flat-sky applications (detailed in Ref.~\cite{shiAKRA2024a}, where it is simply called 'AKRA'), now extended for high-resolution mass mapping on the sphere. AKRA 2.0 includes spin-weighted spherical harmonic transforms, enabling precise mass mapping on the sphere. We refer to this sphere analysis of the algorithm as AKRA-sphere.

Achieving high resolution in full-sky mass mapping also presents a significant challenge.
Current methods, particularly those that utilize spherical harmonic transformations, face limitations when scaled to higher resolutions and larger parameter spaces. According to recent studies \cite{Fiedorowicz2022b}, most full-sky algorithms operate effectively at coarser angular resolutions, typically greater than 10 arcmin \cite {Boruah2024}.
To address computational challenges at high resolutions, AKRA 2.0 incorporates a novel scale-splitting framework. This framework consists of two parts: AKRA-sphere and AKRA-flat. AKRA-sphere is focused on reconstructing the convergence field at large scales using smoothed shear maps. Then AKRA-flat is applied to small-scale convergence field reconstruction within patches where the flat-sky approximation is valid. The final $\kappa$ map is produced by integrating the outputs from both AKRA-sphere and AKRA-flat.

The structure of the paper is organized as follows. In Sec. \ref{sec:method}, we present the derivation and workflow of AKRA 2.0 algorithm. In Sec.~\ref{sec:simulation}, we validate the performance of AKRA 2.0 using simulated shear catalogs incorporating the mask from DESI imaging surveys DR8 and CSST survey forecast. In Sec.~\ref{sec:discussion}, we conclude the results and discuss future applications of AKRA 2.0. Detailed derivations and validations are provided in the appendices.

\section{The AKRA 2.0 Algorithm} \label{sec:method}
\begin{figure*}
    \centering
    \includegraphics[width=0.85\textwidth]{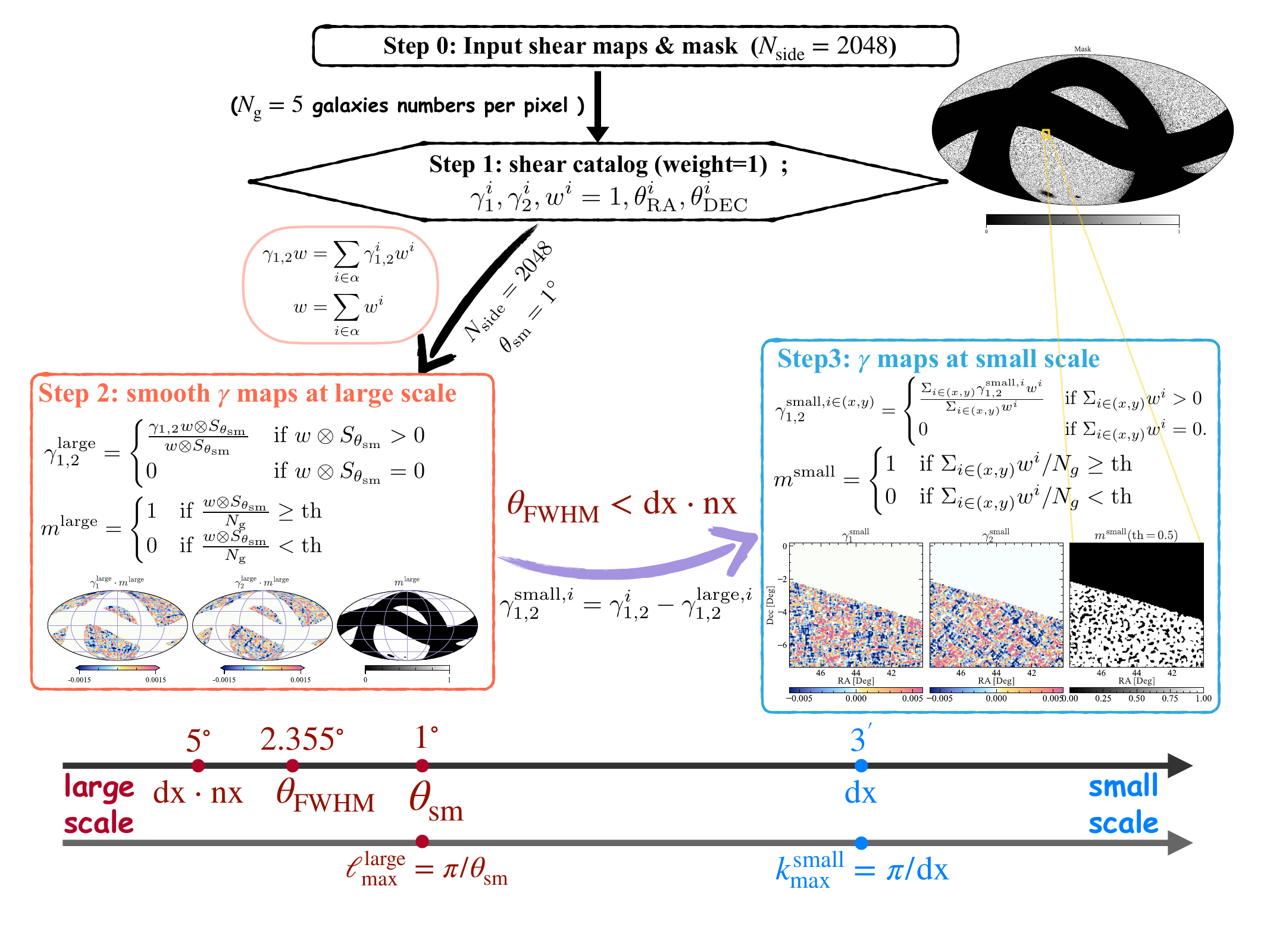}
    \caption{Algorithm flow chart of the AKRA 2.0 algorithm. The process initiates with the yellow hexagon, representing the generation of the observed shear catalog, which includes essential data for each galaxy indexed by $i$: shear components $\gamma_{1,2}^{i}$, weight $w^{i}$, and celestial coordinates ($\theta_{\rm RA}^{i}$, $\theta_{\rm Dec}^{i}$). Subsequently, the workflow is divided into two main analysis steps: Step 1 (red panels) applies the AKRA-sphere algorithm for large-scale analysis, and Step 2 (blue panels) employs the AKRA-flat algorithm for small-scale analysis. The results from both scales are integrated to produce the final convergence map (step 3).}
    \label{fig:framework}
\end{figure*}

The goal of the AKRA 2.0 algorithm is to reconstruct the accurate convergence field 
$\kappa$ from the masked shear catalogs $\gamma$ on the sphere. The AKRA 2.0 algorithm can be  represented as:
\begin{equation}
    \text{\bf AKRA 2.0} = \text{\bf AKRA-sphere} + \text{\bf AKRA-flat}.
    \label{eq:AKRA2.0}
\end{equation} 
For a quick overview, a brief chart summary of the AKRA 2.0 algorithm is presented in Fig.~\ref{fig:framework}.

\subsection{AKRA-sphere: Accurate Kappa Reconstruction Algorithm for masked shear catalog on the sphere}

\subsubsection{Weak Lensing Formalism}
Similar to the decomposition of CMB polarization fields $Q$ and $U$ \cite{huWeakLensing2000a}, the spin-2 shear field $\gamma$ can be expressed in terms of its two components, $\gamma_1$ and $\gamma_2$, as follows:
\begin{equation}
\gamma = \gamma_1 + i \gamma_2  = |\gamma| e^{2 i \phi}
\end{equation}
where $ |\gamma| $ represents the magnitude of the shear and $ \phi $ denotes the phase angle. 
In the harmonics space, then we expand $\gamma$ and its complex conjugate $\gamma^*$ as: 
\begin{equation}
    \begin{aligned}
        \gamma(\hat{\boldsymbol{n}}) &=\sum_{\ell m} a_{2, \ell m }\  { }_{2} Y_{\ell m}(\hat{\boldsymbol{n}}), \\
        \gamma^{*}(\hat{\boldsymbol{n}}) &=\sum_{\ell m} a_{-2, \ell m}\  { }_{-2}Y_{\ell m}(\hat{\boldsymbol{n}}).
    \end{aligned}
    \label{eq:gamma_shear1}
\end{equation}
Here ${ }_{s} Y_{l m}$ denotes the spin-weighted spherical harmonic functions with spin weight $s$. The components $\gamma_1$ and $\gamma_2$ are defined at a specified direction $\hat{\boldsymbol{n}}$ in spherical coordinates $(\hat{e}_{\theta}, \hat{e}_{\phi})$. 
The coefficients $a_{2, \ell m}$ and $a_{-2, \ell m}$ represent the spin-2 spherical harmonic coefficients of the shear field.

In terms of E/B-modes decomposition, Eq.~\ref{eq:gamma_shear1} becomes
\begin{equation}
    \begin{aligned}
        \gamma(\hat{\boldsymbol{n}}) = \sum_{\ell m} \left[ a_{E, \ell m}+i a_{B, \ell m} \right] \ { }_{2} Y_{l m}(\hat{\boldsymbol{n}}),\\
        \gamma^{*}(\hat{\boldsymbol{n}}) = \sum_{\ell m} \left[ a_{E, \ell m}-i a_{B, \ell m} \right] \ { }_{-2} Y_{l m}(\hat{\boldsymbol{n}}).
    \end{aligned}
\end{equation}
where $a_{E, \ell m}$ and $a_{B, \ell m}$ are the spin-0 spherical harmonic coefficients of the E-mode and B-mode polarization fields. They are related to the spin-2 spherical harmonic coefficients $a_{2, \ell m}$ and $a_{-2, \ell m}$ by
\begin{equation}
    \begin{aligned}
        a_{E, \ell m} &= -\frac{1}{2} \left(a_{2, \ell m}+a_{-2, \ell m}\right), \\
        a_{B, \ell m} &= \frac{i}{2} \left(a_{2, \ell m}-a_{-2, \ell m}\right).
    \end{aligned}
    \label{eq:EBmode}
\end{equation}
In the standard cosmological model, the lensing potential $\phi$ only produces the weak gravitational lensing E modes, with $a_{B,\ell m}=0$. Therefore we have 
\begin{equation}
    \begin{aligned}
        a_{E, \ell m} &= -a_{2, \ell m} = -a_{-2, \ell m}.
    \end{aligned}
    \label{eq:Emode_unmasked}
\end{equation}
The relationship among the coefficients $\phi_{\ell m}$, $a_{2,\ell m}$, and $\kappa_{\ell m}$ can be expressed as:
\begin{equation}
    \begin{aligned}
        a_{2, \ell m} &= -a_{E, \ell m} = \frac{1}{2} \sqrt{\frac{(\ell+2)!}{(\ell-2)!}} \phi_{\ell m}, \\
        \kappa_{l m} &= -\frac{\ell(\ell+1)}{2} \phi_{\ell m}, \\
        a_{E, \ell m} &= \sqrt{\frac{(\ell-1)(\ell+2)}{\ell(\ell+1)}} \kappa_{\ell m}.\\
    \end{aligned} 
    \label{eq:kappa_true} 
\end{equation}
Here $\phi_{l m}$ is the coefficients of the lensing potential , and $\kappa_{l m}$ is coefficients of the lensing convergence. The harmonic space relationships in Eq.~\ref{eq:Emode_unmasked} and Eq.~\ref{eq:kappa_true} allow us to transform an unmasked shear coefficients $a_{\pm 2, \ell m}$ into convergence coefficients $\kappa_{\ell m}$ or lensing potential $\phi_{l m}$. This process is performed by the KS algorithm.

\subsubsection{AKRA-sphere algorithm} \label{subsubsec:akra_sphere}

We define a mask function $m(\hat{\boldsymbol{n}})$ in real space, which is 1/0 for the observed/masked region. Then the observed shear field can be expressed as:
\begin{equation} 
    \begin{aligned}
    \gamma^{\m}(\hat{\boldsymbol{n}}) &= m(\hat{\boldsymbol{n}}) \gamma(\hat{\boldsymbol{n}}),\\
    {\gamma^{\m}}^{*}(\hat{\boldsymbol{n}}) &= m(\hat{\boldsymbol{n}}) \gamma^{*}(\hat{\boldsymbol{n}}).
    \end{aligned}
    \label{eq:masked_shear}
\end{equation}
where $\gamma^{\m}(\hat{\boldsymbol{n}})$ and ${\gamma^{\m}}^{*}(\hat{\boldsymbol{n}}) $ are the masked shear field. 
In spherical harmonic space, we have 
\begin{equation}
    \begin{aligned}
        \gamma^{\m}(\hat{\boldsymbol{n}}) &= \sum_{\ell m} a_{2, \ell m}^{\m}\  { }_{2} Y_{l m}(\hat{\boldsymbol{n}}),\\
        {\gamma^{\m}}^{*}(\hat{\boldsymbol{n}}) &= \sum_{\ell m} a_{-2, \ell m}^{\m} \ { }_{-2} Y_{l m}(\hat{\boldsymbol{n}}).
    \end{aligned}
\end{equation}

\begin{comment}
The coefficients $ a_{2, \ell m}^m $ and $ a_{-2, \ell m}^m $ can be computed using:
\begin{equation}
    \begin{aligned}
        a_{2, \ell m}^{\m} &= \int d \Omega \gamma^{\m}(\hat{\boldsymbol{n}})\  { }_{2} Y_{l m}^{*}(\hat{\boldsymbol{n}}),\\
        a_{-2, \ell m}^{\m} &= \int d \Omega {\gamma^{\m}}^{*}(\hat{\boldsymbol{n}}) \ { }_{-2} Y_{l m}^{*}(\hat{\boldsymbol{n}}).
    \end{aligned}
\end{equation}

Then substitue the Eq. \ref{eq:masked_shear} into the above equation, we have:
\begin{equation}
    \begin{aligned}
        a_{2, \ell m}^{\m} &= \int d \Omega m(\hat{\boldsymbol{n}}) \gamma(\hat{\boldsymbol{n}})\  { }_{2} Y_{\ell m}^{*}(\hat{\boldsymbol{n}}),\\
        a_{-2, \ell m}^{\m} &= \int d \Omega m(\hat{\boldsymbol{n}}) \gamma^{*}(\hat{\boldsymbol{n}})\  { }_{-2} Y_{\ell m}^{*}(\hat{\boldsymbol{n}}).
    \end{aligned}
\label{eq:gamma_mask_coeff}
    % a_{2, \ell_1 m_1}^{\m} = \int d \Omega m(\hat{\boldsymbol{n}}) \gamma(\hat{\boldsymbol{n}}) { }_{2} Y_{\ell_1 m_1}^{*}(\hat{\boldsymbol{n}}).
\end{equation}

In the spherical harmonic space, the observed shear field can be expressed as a convolution of 3j-symbols, as shown in Eq. \ref{eq:double_product_3j}. It is worth noting that the mask is typically constant in real observations, thus it can also be considered as a transfer matrix as described in Eq. \ref{eq:double_product_transfer}. More detailed derivation can be found in Appendix \ref{subsec:appendix_convoltion}. Then the $ a_{2, \ell m}^{\m} $ and $ a_{-2, \ell m}^{\m} $ can be expressed as: 
\end{comment}

The spherical harmonic coefficients $ a_{\pm 2, \ell m}^{\m} $ of the masked map are the unmasked spherical harmonic coefficients convoluted with the mask function, which can be expressed as: 
\begin{equation}
    \begin{aligned}
        a_{\pm 2, \ell_1 m_1}^{\m} &= { }_{\pm 2}\mathbf{M}_{\ell_1 m_1 \ell m}\  a_{\pm 2, \ell m},\\
    \end{aligned}
    \label{eq:gamma_obs2}
\end{equation}
or in the matrix form:
\begin{equation}
    \left[\begin{array}{l}a_{2, \ell_1 m_1}^{\m}\\ a_{-2, \ell_1 m_1}^{\m}\end{array}\right]=
    \left[\begin{array}{c}{ }_{2}\mathbf{M}_{\ell_1 m_1 \ell m} \\ { }_{-2}\mathbf{M}_{\ell_1 m_1 \ell m}\end{array}\right] \cdot -a_{E, \ell m},
    \label{eq:gamma_obs3}
\end{equation}
where the coefficients of masked shear field $ a_{\pm 2, \ell_1 m_1}^{m} $ have size of $ (\ell_{\text{max}} + 1)^2 $ for a given $ \ell_{\text{max}} $. The matrix $ { }_{\pm 2} \mathbf{M}_{\ell_1 m_1 \ell m} $ is the convolution kernel with the shape of $ \left((\ell_{\text{max}} + 1)^2, (\ell_{\text{max}} + 1)^2 \right) $. As noted in Eq. \ref{eq:Emode_unmasked}, we have $ a_{2, \ell m} = a_{-2, \ell m} = -a_{E, \ell m} $, indicating that the B-mode coefficients are zero in the absense of mask.

\begin{comment}
Then we rewrite the above equation as matrix form:
\begin{equation}
    \left[\begin{array}{l}a_{2, \ell_1 m_1}^{\m}\\ a_{-2, \ell_1 m_1}^{\m}\end{array}\right]=
    \left[\begin{array}{c}{ }_{2}\mathbf{M}_{\ell_1 m_1 \ell m} \\ { }_{-2}\mathbf{M}_{\ell_1 m_1 \ell m}\end{array}\right] \cdot -a_{E, \ell m},
    \label{eq:gamma_obs3}
\end{equation}
\end{comment}

Same as the AKRA-flat algorithm \cite{shiAKRA2024a}, the masked shear maps are linearly related to the convergence field in spherical harmonic space. By grouping the two components of the shear field, we construct the vector $ \boldsymbol{y} = [a_{2, \ell_1 m_1}^{m}, a_{-2, \ell_1 m_1}^{m}] $ with length $2(\ell_{\text{max}} + 1)^2$, and similarly we have vector $ \boldsymbol{x} \,=\, [ -a_{E, \ell m} ] $ with length  $ (\ell_{\text{max}} + 1)^2 $. Then we express Eq.~\ref{eq:gamma_obs3} into the general linear equations:
\begin{equation}
    \boldsymbol{y}=\mathbf{A} \boldsymbol{x}+\boldsymbol{n}.
    \label{eq:gamma_linear}
\end{equation}
where the matrix $ \mathbf{A} = \left[\begin{array}{c}{ }_{2}\mathbf{M}_{\ell_1 m_1 \ell m} \\ { }_{-2}\mathbf{M}_{\ell_1 m_1 \ell m}\end{array}\right] $
is of dimension $ \left(2(\ell_{\text{max}} + 1)^2, (\ell_{\text{max}} + 1)^2 \right)$, with each element corresponding to the modes of the true convergence field. 

The optimal estimation \cite{Tegmark1997} of Eq.~\ref{eq:gamma_linear} with minimum variance  is
\begin{equation}
    \hat{\boldsymbol{x}}=\mathbf{D} \mathbf{A}^{\rm{T}} \mathbf{N}^{-1} \boldsymbol{y}.
    \label{eq:kappa_estimator2}
\end{equation}
where $\mathbf{D} = \left(\mathbf{A}^{\rm{T}} \mathbf{N}^{-1} \mathbf{A} \right)^{-1}$, and $ \bf{N} \equiv\left\langle\boldsymbol{n} \boldsymbol{n}^{\rm{T}}\right\rangle$ is the noise covariance. For an ill-posed problem, a small regularization matrix $\mathbf{R}\sim 10^{-4}\mathbf{I}$ could be introduced to stabilize the inversion, and now we have $\mathbf{D} = \left(\mathbf{A}^{\rm{T}} \mathbf{N}^{-1} \mathbf{A} + \mathbf{R} \right)^{-1}$. 
Detailed discussions on the computational strategies will be presented in Appendix \ref{sec:app_calculation}.

As a comparison, the KS algorithm, designed for ideal conditions, presumes $m(\hat{\boldsymbol{n}}) = 1$ to determine $\mathbf{A}$, thereby ignoring the impact of the mask and potentially introducing biases in the reconstructed field. While AKRA-sphere involves solving the equations in Eq.~\ref{eq:gamma_linear} combining the mask transfer matrices for both spin-$\pm 2$, which attempts to seek for the exact solution of convergence map without losing information. Here we only consider a binary 0/1 mask and do not apply corrections to the KS results in the power spectrum analysis, this simplification may lead to overestimated errors in the KS reconstruction. Future studies will include comparisons with more advanced reconstruction methods that are designed to handle complex survey masks and noise properties.

In our current implementation, we use a binary $0/1$ mask mainly for simplicity and computational convenience, particularly in constructing the mask convolution matrix $\mathbf{M}$ in our algorithm. The binary mask simplifies the mathematical formulation and allows us to focus on validating the core aspects of our method without the added complexity of varying mask weights.
Importantly, using a binary mask does not result in significant loss of information in our algorithm. The effects of varying survey depth and other observational factors can be naturally incorporated into the noise covariance matrix $\mathbf{N}$ within our framework. By appropriately modeling $\mathbf{N}$, we can account for spatial variations in data quality and noise levels without explicitly using a continuous mask.

\subsubsection{Large-Scale Analysis in AKRA 2.0} \label{subsubsec:large_scale_sphere}
In large-scale analysis, the initial step involves mapping the measured shear catalog from individual galaxies into their corresponding pixel $\alpha$ on the celestial sphere, expressed as:
\begin{equation}
    \begin{aligned} 
        \gamma_{1,2} w & = \sum_{i \in \alpha} \gamma_{1,2}^{i} w^{i},  \\
        w &= \sum_{i \in \alpha} w^{i}.
    \end{aligned}
\end{equation}
Here $\gamma_{1,2}^{i}$ indicates the shear values for each galaxy $i$, and $w^{i}$ represents the weight factor assigned to these galaxies. 

We apply a Gaussian smoothing kernel to effectively filter out small-scale modes. This step is crucial for concentrating the analysis on large-scale modes, thereby minimizing potential divergences caused by smaller-scale fluctuations.  The selection of the kernel width $\theta_{\rm sm}$,  is pivotal as it dictates the scale of features retained in the analysis. 

The remained large-scale part $\gamma^{\text{large}}_{1,2}$ is expressed as:
% \begin{equation}
%     \gamma^{\text{large}}_{1,2} = 
%     \begin{cases} 
%     \frac{(\gamma_{1,2} w) \otimes S_{\theta_{\rm sm}}}{w \otimes S_{\theta_{\rm sm}}} & \text{if } w \otimes S_{\theta_{\rm sm}} > 0 ,\\ 
%     0 & \text{if } w \otimes S_{\theta_{\rm sm}} = 0 .
%     \end{cases}
%     \label{eq:smoothing}
% \end{equation}

\begin{equation}
    \gamma^{\text{large}}_{1,2} = 
    \frac{(\gamma_{1,2} w) \otimes S_{\theta_{\rm sm}}}{w \otimes S_{\theta_{\rm sm}}} 
    \label{eq:smoothing}
\end{equation}

Here, $\otimes$ denotes the convolution, and $S_{\theta_{\rm sm}}$ is the Gaussian smoothing kernel applied at a scale of $\theta_{\rm sm}$. 
The smoothing kernel sets the upper boundary of the multiple on the sphere, $\ell_{\rm{max}}^{\text{large}} = \frac{\pi}{\theta_{\rm sm}}$, and it has a full-width-at-half-maximum (FWHM) determined by $\theta_{\rm sm}$ size:
\begin{equation}
   R_{\rm FWHM} = 2\sqrt{2\ln 2} \theta_{\rm sm}.
\end{equation}

The mask at large scale $m^{\text{large}}$ is generated to focus the analysis on areas with robust data coverage. This mask is defined as:
\begin{equation}
    m^{\text{large}} = 
    \begin{cases}
    1 & \text{if } \frac{w \otimes S_{\theta_{\rm sm}}}{N_{\rm g}^{\text{pixel}}} \geq \text{\it th}, \\ 
    0 & \text{if } \frac{w \otimes S_{\theta_{\rm sm}}}{N_{\rm g}^{\text{pixel}}} < \text{\it th},
    \end{cases}
    \label{eq:mask_large}
\end{equation}
$N_{\rm g}^{\text{pixel}}$ is the average number of galaxies per pixel, and $\text{\it th}$ is the threshold that distinguishes pixels with sufficient galaxies for accurate shear analysis.

Utilizing the large-scale shear maps $\gamma^{\text{large}}_{1,2}$ and the defined mask $m^{\text{large}}$, we employ AKRA-sphere to reconstruct the large-scale convergence field $\kappa^{\text{large}}$.

\subsection{AKRA-flat: Small Scale Analysis} \label{subsec:small_scale_flat}
In the flat-sky limit, a similar relationship in Eq.~\ref{eq:gamma_linear} between the masked shear field and the convergence field has been derived in AKRA-flat algorithm (see Ref.~\cite{shiAKRA2024a} for details).

To effectively manage computational cost while maintaining high precision in our analysis, 
we partition the entire celestial sphere into several manageable flat-sky regions (denoted as $R_1, R_2, \ldots, R_{i\text{\it th}}$)\footnote{The flat sky approximation effectively applies to regions typically smaller than approximately $20^2 \deg^2$. By ``manageable flat-sky regions'', we mean that we divide the celestial sphere into small patches where the flat-sky approximation is valid. These regions are sufficiently small to minimize distortions due to the curvature of the sphere. The regions are designed to slightly overlap to ensure continuity and to mitigate edge effects."}.
For each flat-sky region, we paint $N$ uniform grid with equal grid spacing $H_x=H_y=H$. This segmentation aslo facilitates the use of the flat-sky approximation necessary for AKRA-flat analysis.

We provide a detailed account of the transformation process from spherical to flat-sky coordinates in Appendix \ref{subsec:transform}, a critical step that is validated under the flat-sky approximation with periodic boundary conditions. Visual results of this transformation can be viewed in Fig.~\ref{fig:sphere_to_flat}\footnote{Angular corrections applied to spin-2 shear fields are not trivial, requiring precise alignments with the Cartesian coordinate system. Detailed methodologies and corrections are elaborated in Appendix \ref{subsec:transform}.}. Applying the KS method to shear maps within flat-sky regions, we generate $\kappa$ maps and compare them with the original convergence maps extracted from the spherical field, as shown in Panel (h) of Fig.~\ref{fig:sphere_to_flat}. The high accuracy of the reconstructed $\kappa$ map underscores the effectiveness of the transformation process. However, significant discrepancies are observed at the margins of the map, attributed to the influence of periodic boundary conditions. To address these effects, AKRA-flat specifically improves the treatment of edge areas in the mask by reducing boundary values to zero\footnote{The treatment of boundary effect should be taken carefully in flat-sky mass mapping due to the assumption of periodic boundary conditions. 
We found that these effects are particularly sensitive to the high $\ell$ modes. Therefore, another effective method to reduce the discrepancies caused by the periodic boundary effect is to eliminate the high $\ell$ modes in mass mapping.}.

Based on the successful transformation from spherical to flat-sky coordinates, we proceed to the small-scale analysis using the AKRA-flat algorithm.
Following the extraction of the large-scale shear field $\gamma^{\text{large}}_{1,2}$, we proceed to obtain small-scale shear catalog by subtracting large-scale components from the original shear catalog $\gamma_{1,2}^i$. 
The small-scale shear components for each galaxy, $\gamma^{\text{small},i}_{1,2}$, are calculated as:
\begin{equation} 
    \gamma^{\text{small},i}_{1,2} = \gamma_{1,2}^i - \gamma^{\text{large},i}_{1,2}. 
    \label{eq:small_scale_g}
\end{equation}
where $\gamma^{\text{large},i}_{1,2}$ represents the interpolated large-scale shear values at the positions of individual galaxies from the large-scale shear field $\gamma^{\text{large}}_{1,2}$.

Transitioning to a flat-sky coordinate system introduces complexities due to potential interpolation errors, particularly in the presence of a mask. To mitigate these issues, we utilize mass assignment techniques commonly applied in galaxy survey analyses. Specifically, the Nearest Grid Point (NGP) method is used to allocate the measured shear of each galaxy directly to the closest grid cell, thus minimizing potential interpolation inaccuracies. The one-dimensional assignment function for the NGP method \cite{Jing2005} is:
\begin{equation}
    W_{p=1}(s)=\left\{\begin{array}{ll}1 & \text { for }|s|<\frac{1}{2} \\ 0 & \text { otherwise }\end{array}\right. (\text{NGP}),
\end{equation}
where $s = \frac{x^{G}_i - x}{H}$ is the normalized distance from a galaxy's position $x^{G}_i$ to the nearest grid cell center $x$.

Within each flat-sky region, the small-scale shear field, $\gamma^{\text{small}}_{1,2}$ is computed using a two-dimensional implementation of the NGP method:
\begin{equation}
    \gamma^{\text{small}}_{1,2} = 
    \begin{cases} 
    \frac{\Sigma_{i \in (x,y)} \gamma_{1,2}^{{\rm small},i} w^i}{\Sigma_{i \in (x,y)} w^i} & \text{if } \Sigma_{i \in (x,y)} w^i > 0, \\ 
    0 & \text{if } \Sigma_{i \in (x,y)} w^i = 0 .
    \end{cases}
    \label{eq:small_scale_gmap}
\end{equation}
Here the summation is performed over all galaxies $i$ located within the grid cell $(x,y)$.
Simultaneously, a mask map $m^{\text{small}}$ is computed to identify reliable cells for further scientific analysis. This mask is defined as:
\begin{equation}
    m^{\text{small}} = 
    \begin{cases}
    1 & \text{if } \frac{\Sigma_{i \in (x,y)} w^i}{N_{\rm g}^{\text{grid}}} \geq \text{\it th}, \\ 
    0 & \text{if } \frac{\Sigma_{i \in (x,y)} w^i}{N_{\rm g}^{\text{grid}}} < \text{\it th}.
    \end{cases}
    \label{eq:mask_small}
\end{equation}
$N_{\rm g}^{\text{grid}}$ denotes the average number of galaxies per grid cell.
The threshold value is set to ensure that only cells with sufficient galaxy counts are analyzed, enhancing the reliability of our small-scale convergence maps.

When applying Eq.~\ref{eq:small_scale_g} and Eq.~\ref{eq:small_scale_gmap} , we obtain the small-scale shear field $\gamma^{\text{small}}_{1,2}$ in each region from the shear catalog. Each region's shear map $ \gamma^{\text{small}}_{1,2}$ has overlapped boundaries with neighboring regions to avoid loss of information.

To capture small-scale features and prevent signal aliasing, the maximum $k$-mode is limited by the Nyquist frequency, given by $k_{\text{Nyq}} = \pi / H$. To effective capture of all scale modes, the grid size in small-scale analysis must be selected based on the FWHM of the smoothing kernel used in large-scale analysis, denoted as $R_{\rm FWHM}$. This relationship is given by:
\begin{equation}
    R_{\rm FWHM} < H \cdot N.
\end{equation}
Using the AKRA-flat algorithm, we reconstruct the small-scale convergence maps for each flat-sky region, with each region processed independently. These individual maps are then merged to form the final small-scale convergence map.

The composite map is influenced by the window function due to the mass assignment method. The window functions in Fourier space, associated with each mass assignment scheme, are represented by sinc functions \cite{Jing2005}:
\begin{equation}
    \tilde{W}_{p}(\boldsymbol{k})=\left[\operatorname{sinc}\left(\frac{k_x H_x}{2}\right) \operatorname{sinc}\left(\frac{k_y H_y}{2}\right) \right]^{p} ,
    \label{eq:window_function}
\end{equation}
where $k_x$ and $k_y$ are the wave numbers. The power $p$ is the power of the window function, determined by the mass assignment method used, with the NGP method corresponding to $p=1$.
To avoid duplication or loss of information in the overlapped areas, we compute the averaged convergence maps $\kappa^{\text{small}}$ by averaging the reconstructed values $\kappa^{\text{small},i}$ from all regions covering the same sky area.
By deconvolving this window function from the synthesized small-scale maps from each region, we obtain the small scale convergence map, $\kappa^{\text{small}}$.

\subsection{AKRA 2.0: Combining Large- and Small-Scale Convergence Maps} \label{subsec:combine}
The final convergence map $\kappa$, combines contributions from two different scales:
\begin{equation}
    \kappa^{\text{rec}} = \kappa^{\text{large}} + \kappa^{\text{small}}.
\end{equation}
This formulation aligns with Eq.~\ref{eq:AKRA2.0} and utilizes the $\kappa^{\text{large}}$ and $\kappa^{\text{small}}$ maps, derived from the AKRA-sphere and AKRA-flat algorithms, respectively. 
To ensure that each scale contributes appropriately, we carefully handled the overlapped modes between the large-scale and small-scale maps
\footnote{In the small-scale convergence result $\kappa^{\text{small}}$, we set $\kappa^{\text{small}}(\ell) = 0$ for $\ell < \ell_{\text{FWHM}}$, where $\ell_{\text{FWHM}}$ corresponds to the FWHM of the smoothing kernel at large scale. This is because the small-scale reconstruction can have leakage of power from high $\ell$ modes to low $\ell$ modes due to the finite size of the flat-sky regions and the periodic boundary conditions assumed. By setting these low-$\ell$ modes to zero, we ensure that any potential leakage into the large scales is eliminated.\\
\indent Since $\kappa^{\text{small}}(\ell) = 0$ for $\ell < \ell_{\text{FWHM}}$, the large-scale map must provide the full contribution at these scales. However, $\kappa^{\text{large}}$ is smoothed due to the application of the smoothing kernel, which should be de-convolved in final convergence map.}.

The scale-splitting approach of AKRA 2.0 effectively balances computational demands with the precision required for high-resolution mapping. In this work, we concentrate on the AKRA 2.0 algorithm and its scale-splitting framework. Based on this algorithm, the subsequent section will present the validation of AKRA 2.0 using various masks and compare its performance with the KS method.

\section{Validation of AKRA 2.0}
\label{sec:simulation}

We validate AKRA 2.0 under various simulated conditions. This includes testing its individual components, AKRA-sphere and AKRA-flat, as well as the integrated performance of the entire AKRA 2.0 algrithm.

Key statistical measures employed for performance comparison include: the slope of $\kappa^{\rm rec}$ versus $\kappa^{\rm true}$, the Pearson Correlation Coefficient (PCC) $\rho$, the localization measure $L$, the analysis of power spectrum ratios $C_{\ell}^{\text{rec}} / C_{\ell}^{\text{true}}$, and the cross-correlation coefficients $r_{\ell}$. As a reminder, the localization measure $L$ quantifies the extent to which residuals are localized within masked regions, and is defined as:
\begin{equation}
    L \equiv \frac{\sum_{\rm{ipix}} |\Delta_{\rm{ipix}}| m_{\rm{ipix}}}{(\sqrt{2/\pi}\sigma_{\kappa^{\rm true}})(N_{\rm pix}(1-f_{\rm mask}))},
\label{eq:localization}
\end{equation}
where $\Delta_{\rm{ipix}} \equiv \kappa^{\rm rec}_{\rm{ipix}} - \kappa^{\rm true}_{\rm{ipix}}$ represents the residual at each pixel, and $\rm{ipix}$ denotes the pixel index. Tab.~\ref{tab:results} summarizes the performance statistics for the AKRA 2.0 algorithm.

In this work, we generate the Gaussian random field on the sphere as the true convergence using a power spectrum derived from cosmological parameters in the Planck 2018 flat $\Lambda$CDM cosmology \cite{Planck2018}, implemented in \texttt{pyccl} \cite{pyccl}. From the convergence map, we derive the shear fields $\gamma_{1}(\hat{\boldsymbol{n}})$ and $\gamma_{2}(\hat{\boldsymbol{n}})$ using Eq.~\ref{eq:kappa_true} in spherical harmonic space, followed by an inverse spherical harmonic transform to obtain the shear map in real space.

\subsection{Testing AKRA-sphere}
\label{subsec:sphere_performance}
Performance assessments of the AKRA-sphere were focused on solely large-scale analyses. This included testing with binary masks derived from actual observations by the DESI imaging surveys DR8, alongside various simulated mask scenarios. 

We perform the AKRA-sphere test using a pixel resolution of $\approx 28$ arcmin. This resolution corresponds to a setting of $N_{\text{side}} = 128$ and $\ell_{\text{max}} = 195$\footnote{To avoid aliasing effects, we therefore limit our map reconstruction to well-resolved modes with $\ell_{\max} \leq 1.5  N_{\rm{side}}$.  This ensures that the $ a_{\ell m} $ and $C_{\ell} $ computed by the \texttt{HEALPix} are accurate to machine precision without the need for iterations for band-width limited input signal. }.
In this section, we present the validation of DESI imaging surveys DR8 mask. More simulated random mask scenarios will be discussed in Appendix \ref{sec:app_simulation_sphere}.

\begin{figure*}[htbp]
    \centering
    \includegraphics[width=0.95\textwidth]{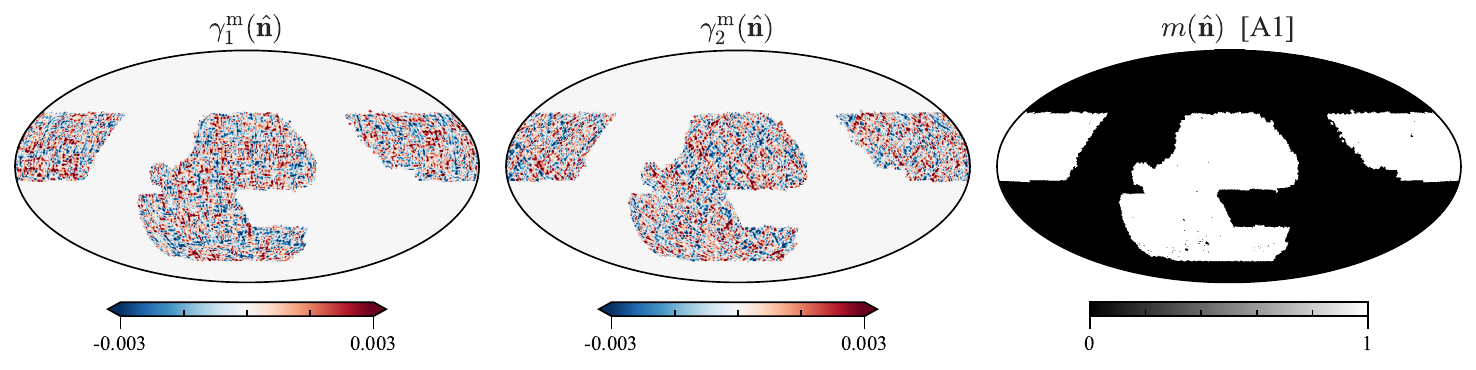}
    \caption{Input data applying DESI imaging surveys DR8 mask. From left to right, the panels show the masked shear field $\gamma_{1}^{m}(\hat{\boldsymbol{n}})$, $\gamma_{2}^{m}(\hat{\boldsymbol{n}})$, and the mask $m(\hat{\boldsymbol{n}})$ (referred to as {\bf A1}), respectively.}
    \label{fig:DESI_g1g2Mask}
\end{figure*}

We generate a binary mask based on the real observations from the DESI imaging surveys DR8 with nside = 1024, utilizing a dataset of 5.18 million galaxies at redshift bin $0.8<z<1.0$, which covering $\sim$ 13,000 deg$^2$. 
The mask is 1 when if are shear galaxies located at the pixel, otherwise it is 0. Then downgrade the resolution to nside = 128. The downgraded mask is 1 when pixel values $\geq 0.5$, otherwise is 0. 
As illustrated in Fig.~\ref{fig:DESI_g1g2Mask}, the input data during the AKRA-sphere simulation is the masked shear field $\gamma_{1}^{m}(\hat{\boldsymbol{n}})$ and $\gamma_{2}^{m}(\hat{\boldsymbol{n}})$, which are generated by applying the mask to the shear field $\gamma_{1}(\hat{\boldsymbol{n}})$ and $\gamma_{2}(\hat{\boldsymbol{n}})$, respectively. 

\begin{figure*}[htbp]
    \subfigure[]{\includegraphics[width=0.7\textwidth]{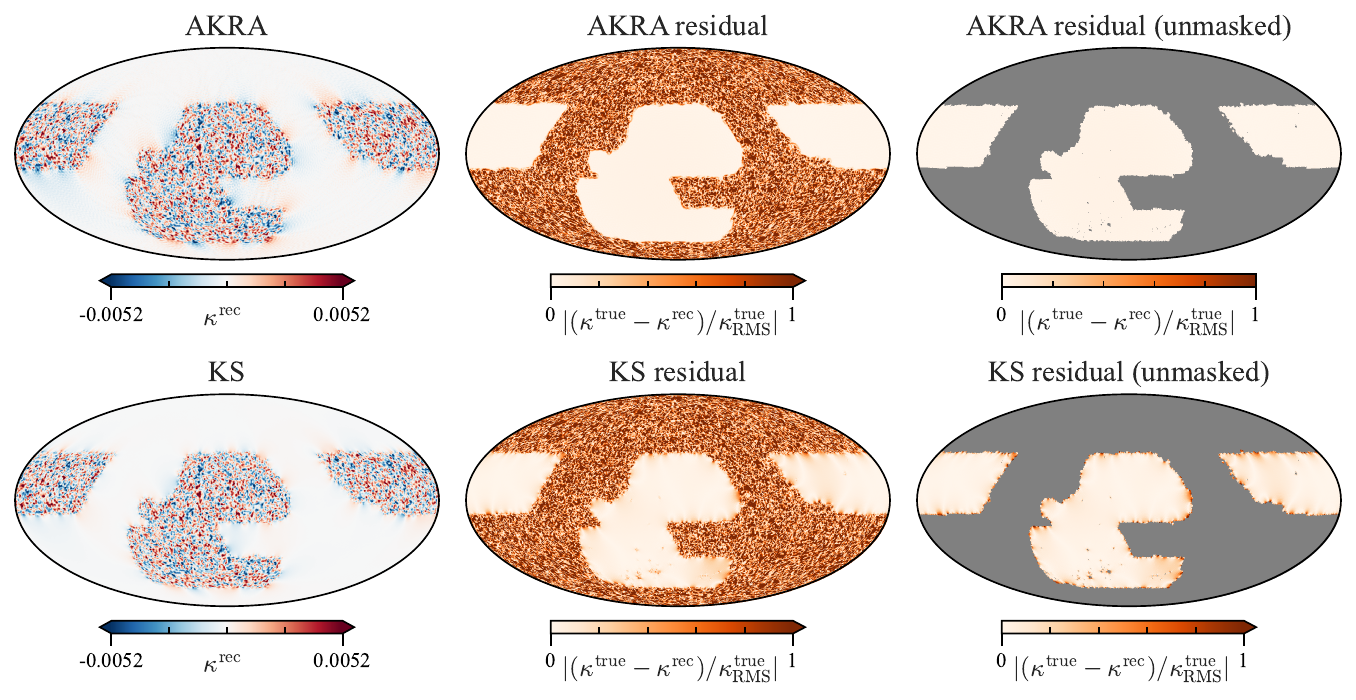}} 
    \subfigure[]{\includegraphics[width=0.28\textwidth]{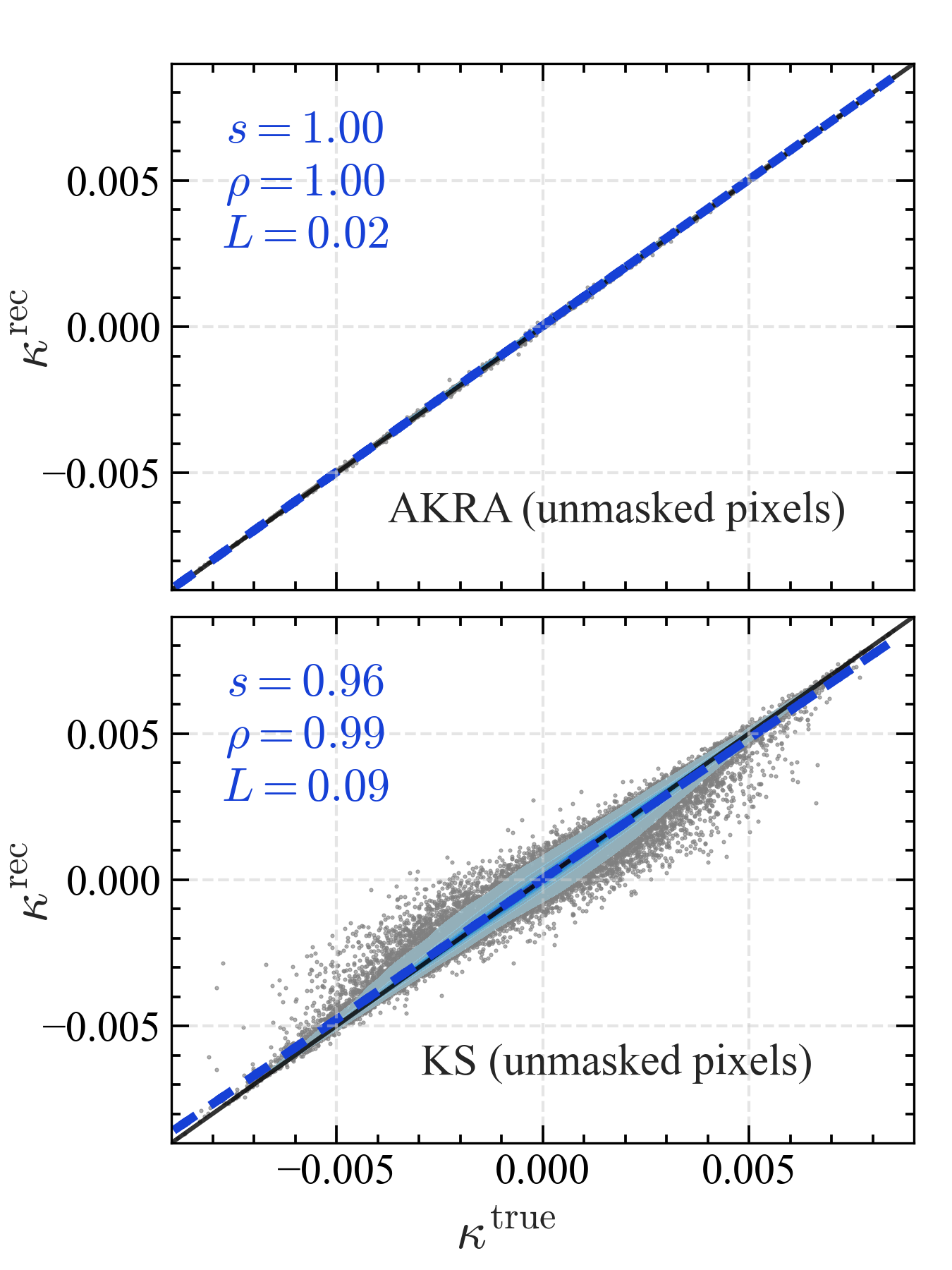}}
    \caption{Results for the DESI imaging surveys DR8 mask using AKRA 2.0 (top row) and the KS direct inversion method (bottom row). (a) Comparison of the results is made with the reconstructed map $\kappa^{\rm{rec}}$ and residual maps normalized by the root mean square (r.m.s.) of the true signal $| \kappa^{\text{true}} - \kappa^{\text{rec}}/\kappa^{\text{true}}_{\text{RMS}}|$. (b) The $\kappa^{\rm rec}$-$\kappa^{\rm true}$ scatter plot for unmasked pixels. The data points are displayed as gray dots, with the x-axis and y-axis denoting the pixel values for the input $\kappa$ map and the reconstructed $\kappa$ map, respectively. The blue dashed line represents the result obtained from fitting a regression model to the pixels from the data points, while the black solid line indicates the ideal result. The slope $s$ of the blue dashed line, PCC $\rho$, and localization measure $L$ are also shown in the panels. } 
    \label{fig:DESI_g1g2Mask-results} 
\end{figure*}

Fig.~\ref{fig:DESI_g1g2Mask-results} illustrates the mass mapping results obtained from AKRA 2.0 (top row) and the KS method (bottom row). 
When comparing the residuals in the unmasked pixels, significant residuals are evident at the edges of the survey footprint for the KS method, with errors approaching the magnitude of the $\kappa$ field we aim to reconstruct. In contrast, AKRA 2.0 demonstrates superior robustness, exhibiting negligible residuals across these areas. 

To evaluate the global quality of the maps in real space, we constructed a $\kappa^{\rm rec}$-$\kappa^{\rm true}$ scatter plot for unmasked pixels, as shown in Fig.~\ref{fig:DESI_g1g2Mask-results} (b). 
The reconstructed $\kappa$ maps produced by AKRA 2.0 are virtually indistinguishable from the original $\kappa$ map, with $s=1.00$ and $\rho=1.00$, demonstrating high accuracy. In contrast, $\kappa$ maps reconstructed using the KS method show noticeable discrepancies, particularly at edge pixels within the mask, resulting in $s=0.96$ and $\rho=0.99$. Furthermore, we defined a localization measure, $L$, to evaluate the extent to which the mask in the shear catalog contaminates unmasked pixels. For AKRA 2.0, $L$ is 0.02, indicating negligible contamination, which is approximately five times lower than that of the KS method, where $L$ is 0.09.

\begin{figure*}[htbp]
    \centering
    \includegraphics[width=0.8\textwidth]{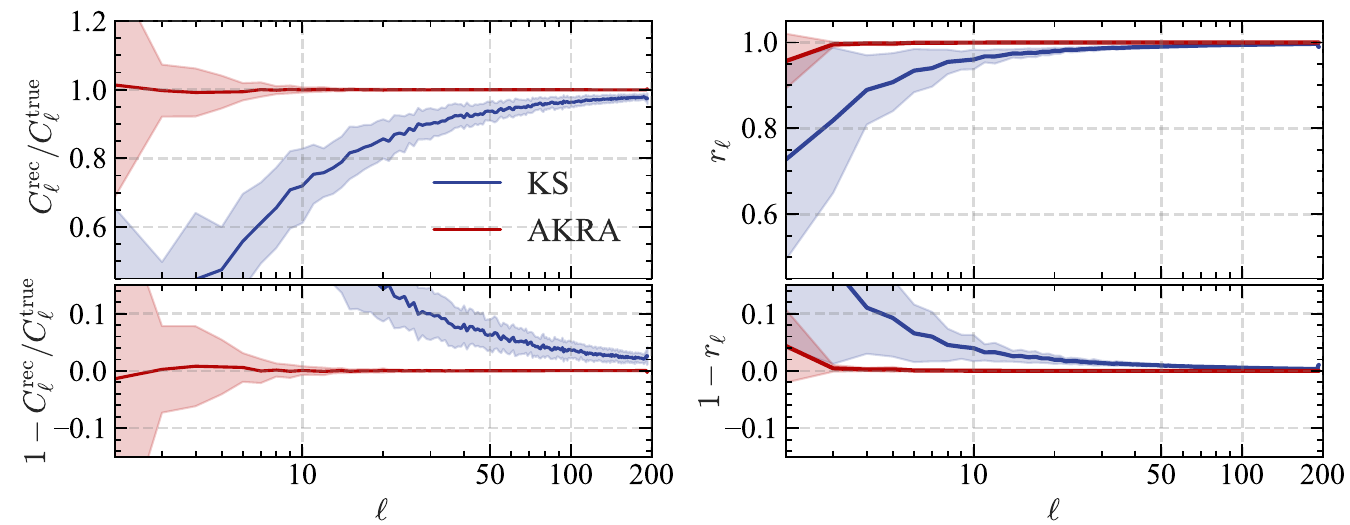}
    \caption{Results for 100 realizations. Power spectrum ratio (left panel) and cross-correlation coefficient (right panel) using the DESI imaging surveys DR8 mask. The blue and red regions represent the 1$\sigma$ confidence interval. The deviations from $1$ are also illustrated in the bottom panels.}
    \label{fig:DESI_rk}
\end{figure*}

The power spectrum ratios and cross-correlation analysis depicted in Fig.~\ref{fig:DESI_rk} further substantiate the superior accuracy of AKRA 2.0. The values of $C_{\ell}^{\text{rec}} / C_{\ell}^{\text{true}}$ and $r_{\ell}$ maintain deviations of less than 1\% from the ideal unity. Additionally, the confidence intervals for AKRA 2.0 are significantly narrower than those for the KS method. These performance statistics illustrate the excellent precision of AKRA 2.0 in reconstructing mass distribution in unmasked pixels.

Simulated random masks were also employed to test the AKRA-sphere algorithm under different percentages of masked pixels and variations in mask sizes. The results confirmed the algorithm's exceptional precision across a range of mask conditions. 
For random mask conditions, the localization measure $L$ for AKRA-sphere is significantly reduced, a factor of ${\mathcal O}(10)$ to ${\mathcal O}(10^2)$ smaller than that of the KS method, as demonstrated in Tab.~\ref{tab:results}. Typically, $L$ in AKRA-sphere for random masks is $\lesssim 10^{-2}$, meaning that contamination from small masked pixels is generally negligible.
Quantitative analysis showed that both the power spectrum ratios, $C_{\ell}^{\text{rec}} / C_{\ell}^{\text{true}}$, and the cross-correlation coefficients, $r_\ell$, maintained deviations less than 1\% from the ideal unity. This performance underscores the AKRA-sphere's capability to accurately process complex survey geometries on the sphere, ensuring reliable reconstructions of mass distribution, as detailed in Appendix \ref{sec:app_simulation_sphere}.

\subsection{Testing AKRA-flat}
\label{subsec:test_flat}
In our previous work \cite{shiAKRA2024a}, we extensively tested the AKRA-flat algorithm under a variety of simulated conditions. These tests included observable masks, random masks with mask fractions ranging from 10\% to 50\%, and masks of varying sizes, all within the context of a flat-sky approximation with periodic boundary conditions. We highlight the exceptional precision achieved with AKRA-flat, which demonstrated $C_{\ell}^{\text{rec}} / C_{\ell}^{\text{true}}$ and $r_\ell$ are both accurate to 1\% or better.
The high level of accuracy achieved by AKRA-flat is essential for its foundational role in small-scale analysis within the AKRA 2.0 framework.

\subsection{Testing AKRA 2.0}
To test AKRA 2.0's performance in high-resolution scenarios, we utilized a simulated CSST-like mask designed to mimic the observational challenges anticipated in the forthcoming CSST survey. 

\begin{figure}[h]
    \centering
    \includegraphics[width=0.6\textwidth]{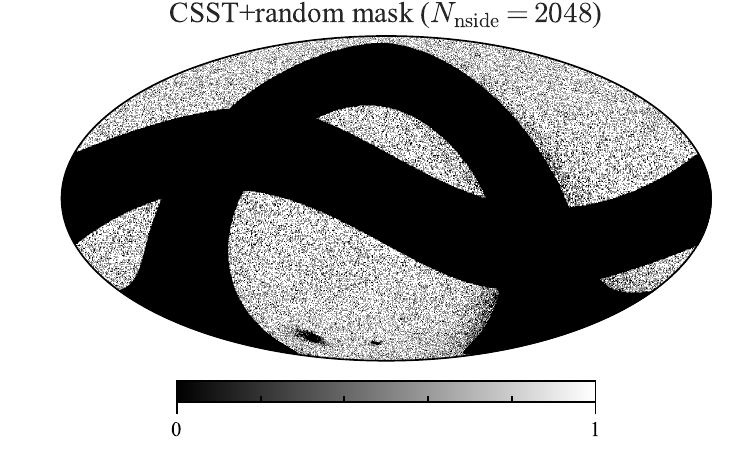}
    \caption{An illustration of the CSST-like mask used in the CSST forcast simulation.   This mask combines the CSST mask derived from cosmic shear forecasts, with a random mask having a 20\% mask fraction. The mask will be employed to evaluate the performance of both the AKRA 2.0 and KS methods.}
    \label{fig:CSST_mask_and_random}
\end{figure}

\begin{figure*}[htbp]
    \centering
    \includegraphics[width=0.95\textwidth]{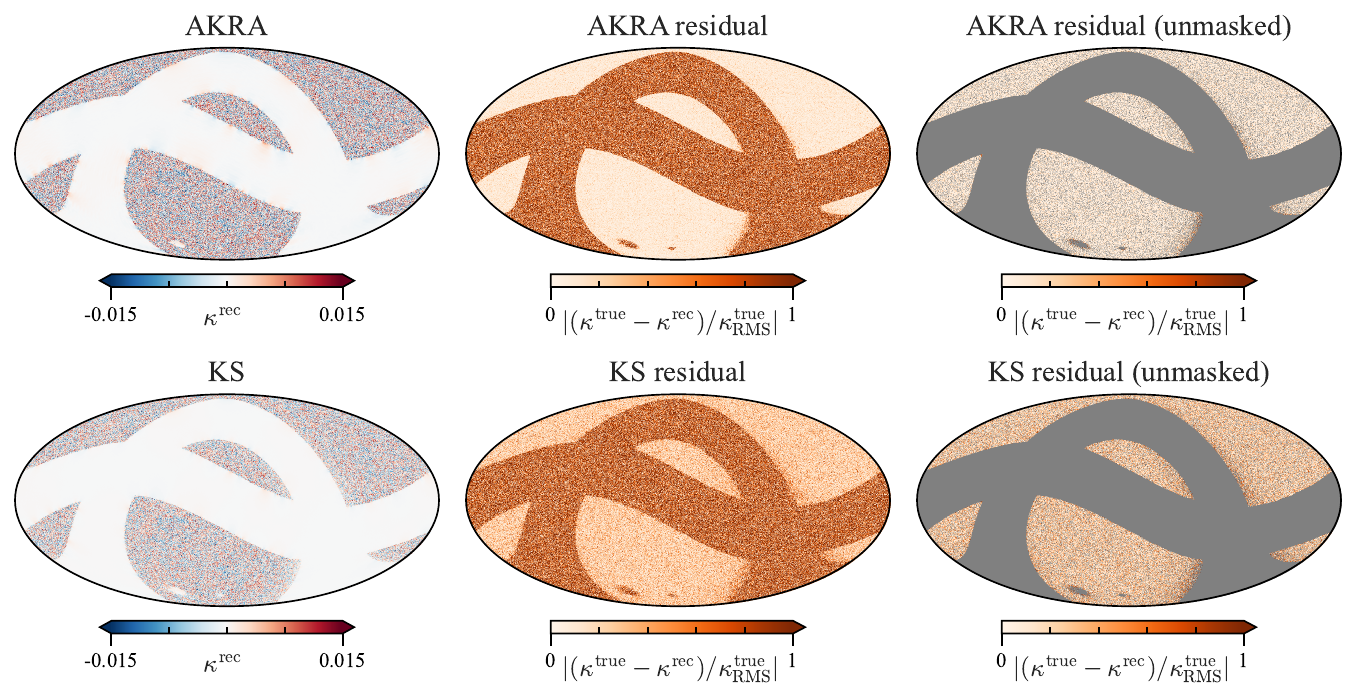}
    \caption{Comparison of reconstructed $\kappa$ map results for the CSST-like mask using AKRA 2.0 (top row) and the KS direct inversion method (bottom row). The residuals from AKRA 2.0 are notably smaller than those from the KS method, indicating a superior fit to the true convergence values. }
    \label{fig:CSST_results}
\end{figure*}

\begin{figure*}
    \centering
    \includegraphics[width=0.85\textwidth]{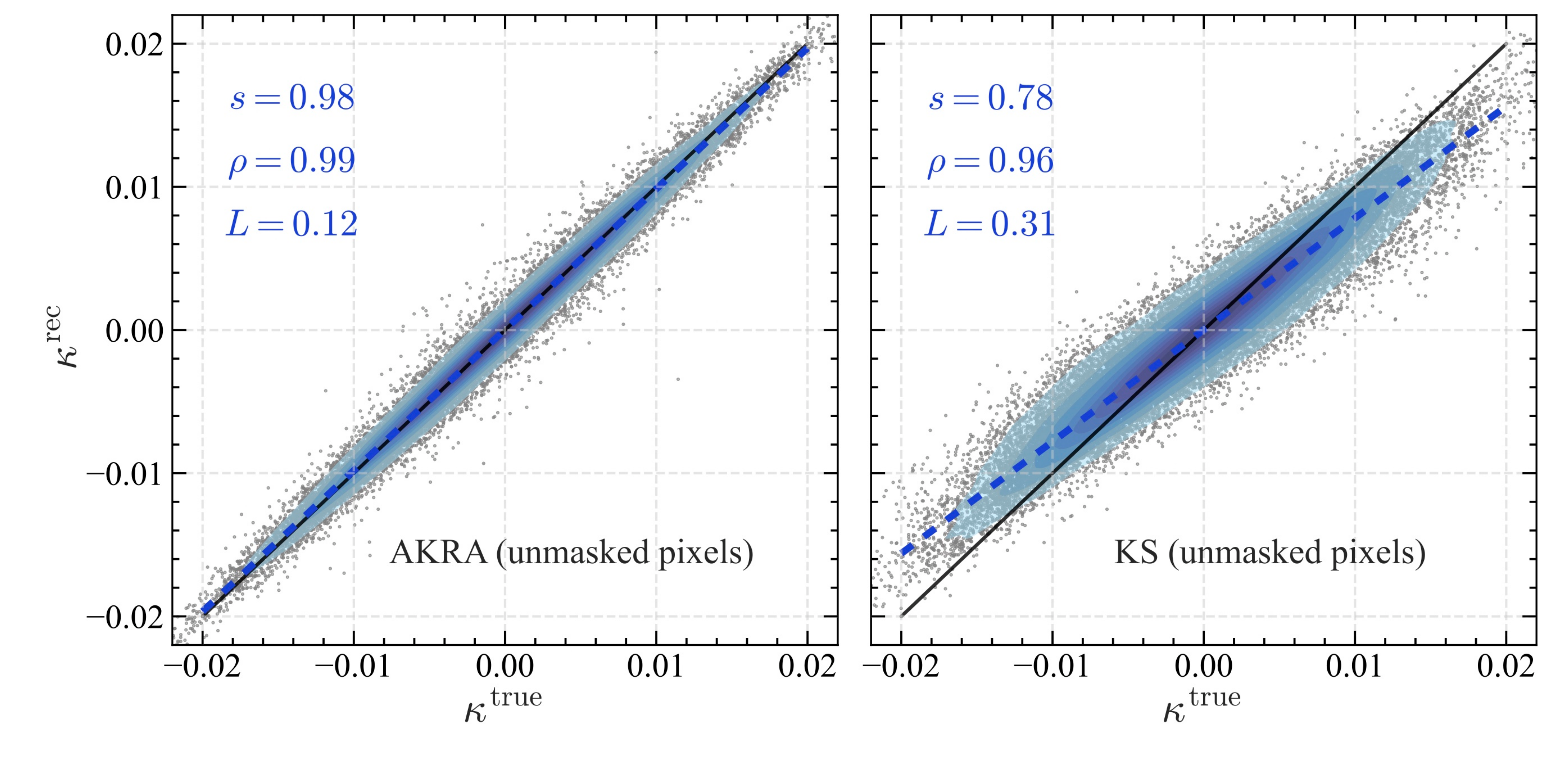}
    \caption{Kernel Density Estimation (KDE) plot of the reconstructed versus true convergence values ($\kappa^{\rm rec}$ vs. $\kappa^{\rm true}$) for the CSST-like mask (in unmasked pixels). This scatter plot includes a random selection of 50,000 pixels from approximately 13,570 $\mathrm{deg}^2$ of unmasked area, out of a total of 16,550,212 pixels at a \texttt{HEALPix} resolution of $N_{\rm side} = 2048$. The ideal relationship, where $\kappa^{\rm rec} = \kappa^{\rm true}$, is depicted by the black solid line with a slope $s = 1.0$. Compared to the KS method, AKRA 2.0 demonstrates superior performance, indicated by a steeper slope closer to the ideal and a higher PCC $\rho$. Furthermore, the localization measure $L$ for AKRA 2.0, at 0.12, reflects more accurate reconstruction than the KS method at 0.31, showcasing the effectiveness of AKRA 2.0 in handling complex mask geometries.}
    \label{fig:CSST_results_kde}
\end{figure*}

\subsubsection{CSST-like mask and simulation setups}
\label{subsec:csst_mask}
We generate the CSST-like mask by integrating the CSST mask, derived from cosmic shear forecasts, with a random mask that has a 20\% mask fraction. The CSST mask is estimated based on the observed distribution of galaxies and the positions of bright stars, as detailed in \cite{Yao2023prr}. To optimize the observational data, we exclude regions within $\pm 19.2^{\circ}$ of the galactic and ecliptic latitudes, as well as areas containing bright sources. Initially, the CSST mask is constructed at a high resolution with $N_{\rm side} = 4096$, which is then downgraded to $N_{\rm side} = 2048$. During this process, we apply a binary thresholding method where the value of the mask is set to 1 if the pixel value exceeds 0.5; otherwise, it is set to 0. This thresholding adjusts the effective sky area covered by the CSST mask to 16,963 $\mathrm{deg}^2$.

Additionally, the random mask is created by randomly assigning zero values to 20\% of the pixels, to simulate the potential observational obstructions by small masks. The integration of these two masks results in a CSST-like mask, covering an effective observational area of 13,570 $\mathrm{deg}^2$. 
The combined mask map is illustrated in Fig.~\ref{fig:CSST_mask_and_random}.

The following simulation setups are chosen to mimic expected observational conditions:
\begin{itemize}
    \item \textit{Resolution:}   The simulations were conducted using \texttt{HEALPix} maps at a resolution parameter of $N_{\text{side}} = 2048$. This choice offers a pixel size of approximately $1.72'$, allowing for high-resolution analysis across the celestial sphere with the maximum multiple moment set to $\ell_{\text{max}} = 3072$.
    \item \textit{Shear Catalog density:}   To mimic the observational density expected in next-generation cosmic surveys, we set the average number of galaxies per pixel at $N_{\text{p}} = 5$ for a total \texttt{HEALPix} resolution of $N_{\text{side}} = 2048$. This setup generated a total catalog of approximately $8.3 \times 10^7$ galaxies, each assigned a weighting factor of $w^{i} = 1$ for simplicity.
    This high density of galaxies per pixel significantly also reduces interpolation errors when transitioning from spherical to flat-sky coordinates\footnote{Details on the mitigation of interpolation errors through sampling are discussed in Ref.~\cite{cunningtonAccurate2024}. This setup serves as a simplified model of what is anticipated in real observational data, particularly in the context of forthcoming Stage IV cosmic surveys where the data density is expected to be significantly higher.}.
\end{itemize}

\subsubsection{AKRA 2.0 analysis}

In the large-scale analysis, the primary focus is on smoothing the data to suppress smaller-scale fluctuations and highlight large scale information. We apply a Gaussian smoothing kernel with $\theta_{\rm sm} = 1^\circ$, corresponding to a FWHM size of $R_{\rm FWHM} = 2\sqrt{2\ln 2} \theta_{\rm sm} \approx 2.35^\circ$.
The maximum multiple moment is restricted to $\ell_{\rm{max}}^{\text{large}} = 180$, determined by the smoothing kernel. The effective FWHM in multiple space is thus $\ell_{\rm FWHM}=76$.
The mask threshold $\text{\it th} = 0.5$ is set to ensure that only sufficiently sampled areas contribute to the analysis.

The small-scale analysis concentrates on extracting small scale information. The sphere is divided into numerous flat-sky regions, each with grid dimensions $N_x = N_y = 150$ and grid spacing $H_x = H_y = 0.05^\circ$, covering approximately $7.5^2 \deg^2$ areas. We position the galaxies on the grid cells by employing NGP mass assignment method. Due to the assumption of periodic boundary conditions, the outer 1/3 of each region's boundary is set to be zero, focusing the analysis on the central 2/3 
(approximately $5^2 \deg^2$). The mask threshold is set to $\text{\it th} = 0.5$ to ensure that only reliable regions are considered. 

\begin{figure}
    \centering
    \includegraphics[width=0.85\textwidth]{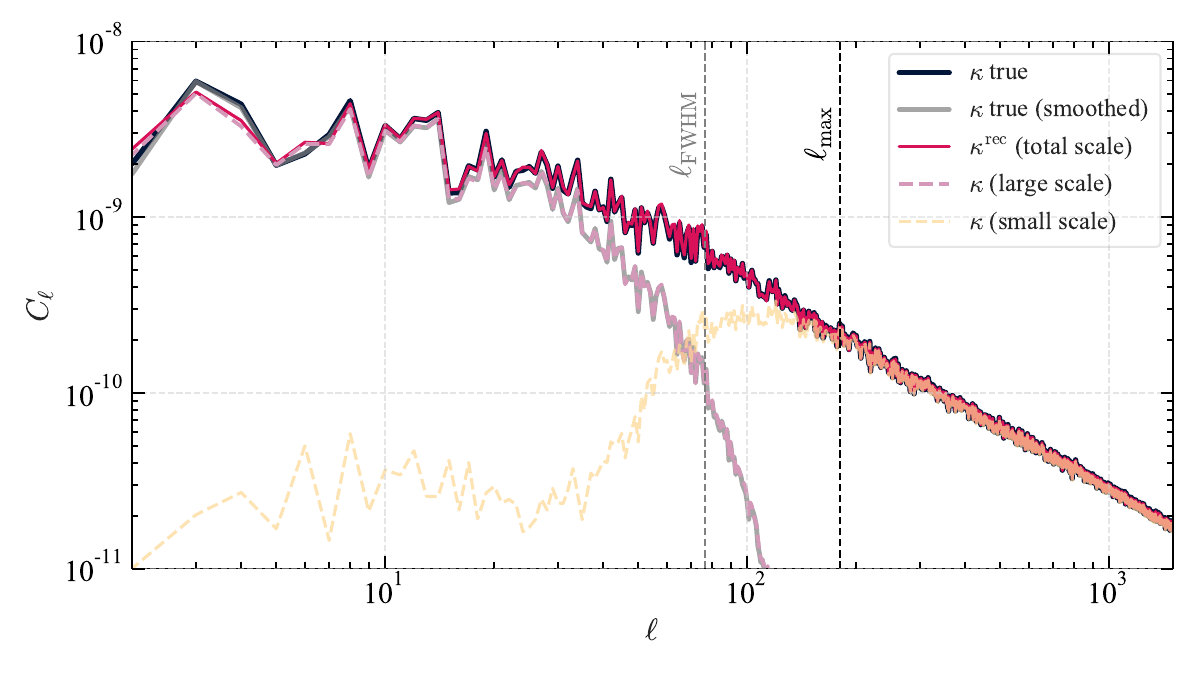}
    \caption{The power spectrum of the true convergence map and the components from our scale-splitting method. We focus on results in unmasked pixels. To prevent information leakage from small to large scales, we filter out large-scale components from the small-scale map (i.e., \(\ell < \ell_{\text{FWHM}}\)). Here the $\ell_{\text{FWHM}}$ is set at $\ell \approx 76$, corresponding to the FWHM of the smoothing kernel $\theta_{\text{sm}} = 1^\circ$ used at large scales.
    For the large scale map, the smoothing kernel should be de-convolved to obtain the true convergence map for $\ell < \ell_{\rm FWHM}$. The final power spectrum of the combined reconstruction $\kappa^{\text{rec}}$ is shown by the red solid line, demonstrating excellent agreement with the true power spectrum.
    Here we focus on the power spectrum up to $\ell = 1500$, which represents half of the Nyquist frequency employed in small-scale analysis.}
    \label{fig:CSST_scale_results}
\end{figure}

The final convergence map $\kappa$ results from synthesis of both large-scale and small-scale maps.
The large-scale map, $\kappa^{\text{large}}$, and the small-scale map, $\kappa^{\text{small}}$, are combined. In Fig.~\ref{fig:CSST_scale_results}, we illustrate how the large-scale and small-scale contributions add up to reconstruct the true convergence power spectrum across all $\ell$. For the small-scale map, deconvolving the NGP window function corrects for the effects of pixelation on the recovered amplitudes, ensuring accurate representation of the high-$\ell$ modes. By setting $\kappa^{\text{small}}(\ell) = 0$ for $\ell < \ell_{\text{FWHM}}$, we prevent leakage of small-scale signals into the large scales. For the large-scale map, we deconvolve the smoothing kernel in $\ell$\-space to correct for its effect on the amplitudes. This ensures that the large-scale modes ($\ell < \ell_{\text{FWHM}}$) are accurately represented.

% To ensure no information leakage from small to large scales, we filter out large-scale components from the small-scale map (i.e., $\ell < \ell_{\rm FWHM}$).

\begin{figure*}[htbp]
    \centering
    \includegraphics[width=0.85\textwidth]{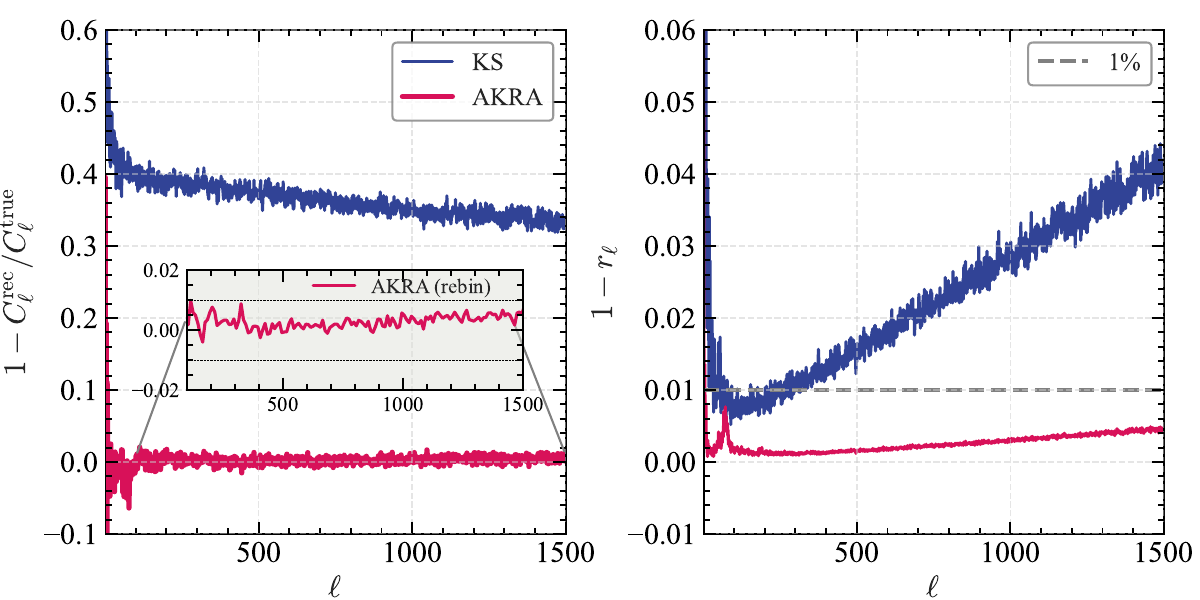}
    \caption{Analysis of the power spectrum ratios (left) and cross-correlation coefficients (right) for the CSST-like mask using AKRA 2.0 and KS methods is presented. The gray panel in the left figure shows averaged results for 10 modes each.
    We focus on results in unmasked pixels, setting masked pixels to zero. Here we do not apply corrections to the KS results in the power spectrum analysis in the current implementation. This simplification may lead to overestimated errors in the KS reconstruction.}
     Notably, AKRA 2.0 shows robust accuracy in power spectrum recovery, eliminating the need for additional corrections.
    \label{fig:CSST_mask_rk}
\end{figure*}

\begin{table}[ht]
    \centering
    \begin{threeparttable}
        \caption{Summary of the AKRA 2.0 performance. Based on the results of $s$, $\rho$, and $L$, it is evident that AKRA 2.0 outperforms the KS method in all cases.}
        \label{tab:results}
        \begin{tabular}{p{2.9cm}p{1.6cm}p{1.6cm}p{1.8cm}}
        \midrule[1pt]
    \textbf{Type} & \textbf{$s$} & \textbf{$\rho$} & \textbf{$L$} \\
    \midrule[1pt]
    \multicolumn{4}{l}{{\bf Testing AKRA-sphere:}} \\
    
    \multicolumn{4}{l}{\textit{DESI imaging surveys DR8 mask} [A1]}\\
    AKRA 2.0  & 1.00 & 1.00 & 1.82$\times 10^{-2}$ \\
    KS  & 0.96 & 0.99 & 9.25$\times 10^{-2}$ \\
    \multicolumn{4}{l}{\textit{Random mask} [B1, B2, B3]\footnotemark[1]}\\ 
    AKRA 2.0 [B1]  & 1.00 & 1.00 & 2.00$\times 10^{-3}$ \\
    AKRA 2.0 [B2]  & 1.00 & 1.00 & 3.61$\times 10^{-3}$ \\
    AKRA 2.0 [B3]  & 1.00 & 1.00 & 6.20$\times 10^{-3}$ \\
    KS [B1]  & 0.91 & 0.98 & 1.83$\times 10^{-1}$ \\
    KS [B2]  & 0.73 & 0.93 & 3.87$\times 10^{-1}$ \\
    KS [B3]  & 0.55 & 0.85 & 5.57$\times 10^{-1}$ \\
    \multicolumn{4}{l}{\textit{Random mask} [C1, C2, C3]\footnotemark[2]}\\ 
    AKRA 2.0 [C1]  & 1.00 & 1.00 & 7.23$\times 10^{-3}$ \\
    AKRA 2.0 [C2]  & 1.00 & 1.00 & 3.33$\times 10^{-3}$ \\
    AKRA 2.0 [C3]  & 1.00 & 1.00 & 4.54$\times 10^{-3}$ \\
    KS [C1]  & 0.81 & 0.95 & 3.03$\times 10^{-1}$ \\
    KS [C2]  & 0.72 & 0.92 & 3.88$\times 10^{-1}$ \\
    KS [C3]  & 0.64 & 0.89 & 4.74$\times 10^{-1}$ \\
    \midrule[1pt]    
    \multicolumn{4}{l}{{\bf Testing AKRA-flat:}} \\
    \multicolumn{4}{l}{Detailed results can be found in AKRA-flat work \cite{shiAKRA2024a}.} \\
    \midrule[1pt] 
    \multicolumn{4}{l}{{\bf Testing AKRA 2.0:}} \\
    \multicolumn{4}{l}{\textit{CSST-like mask}}\\
    AKRA 2.0\footnotemark[3]  & 0.98 & 0.99 & 1.21$\times 10^{-1}$  \\
    KS  & 0.78 & 0.96 & 3.10$\times 10^{-1}$  \\

    \midrule[1pt]    

    \end{tabular}
        \begin{tablenotes}
            \footnotesize
            \item[1] B1, B2, and B3 denote random masks with 10\%, 30\%, and 50\% mask fractions, respectively. 
            \item[2] C1, C2, and C3 represent random masks fixed at a 40\% mask fraction but with varying mask sizes.
            \item[3] The results for the CSST-like mask are also influenced by the resampling errors inherent in the transformation process used by AKRA 2.0, but they still outperform those obtained using the KS method.
        \end{tablenotes}
    \end{threeparttable}
\end{table}

\subsubsection{Mass Mapping Results for the CSST-like Mask}
The reconstructed $\kappa$ maps, illustrated in Fig.~\ref{fig:CSST_results}, reveal that AKRA 2.0 produces notably smaller residuals compared to the KS method. The residuals for unmasked pixels in AKRA 2.0 exhibit minor deviations from zero, primarily due to resampling errors inherent in the transformation process used by AKRA 2.0. 
The $\kappa^{\rm rec}$ versus $\kappa^{\rm true}$ scatter plot in Fig.~\ref{fig:CSST_results_kde} also quantifies the accuracy of AKRA 2.0. AKRA 2.0 maintains a steeper slope $s$ and higher PCC $\rho$ of 0.98 and 0.99, close to 1.0. 
This high accuracy is contrasted against the results from the KS method, which shows a marked reduction in accuracy with $s = 0.78$ and $\rho = 0.96$. 
Additionally, AKRA 2.0's localization measure ($L$) of 0.12 is significantly lower than the KS method's 0.31, indicating less contamination from masked pixels. These statistics are summarized in Tab.~\ref{tab:results}.

Moreover, the analysis of power spectrum ratios and cross-correlation coefficients, as presented in Fig.~\ref{fig:CSST_mask_rk}, further underscores the superior performance of AKRA 2.0. Both metrics deviate from unity by less than 1\%, which highlights AKRA 2.0's precise recovery of the underlying mass field's power spectrum. This precision in power spectrum can avoid the need for further corrections typically necessary when using the KS methods. 

It is important to note that in our implementation of the KS method, we assumed $m(\hat{n})$ = 0/1 and did not apply any corrections for the mask in the power spectrum analysis. Our comparisons, therefore, serve to highlight the improvements of AKRA 2.0 over the basic KS method under these assumptions. A more sophisticated implementation of the KS algorithm could mitigate some of these errors by incorporating the mask into the reconstruction and correcting the power spectrum. However, our focus here is to provide a baseline comparison using the foundational KS method.

The superior performance of AKRA 2.0 thus positions it as an valuable tool in the analysis of CSST data, as the precision of mass mapping directly influences the accuracy of inferred cosmological parameters. 
Moreover, these results not only validate the effectiveness of AKRA 2.0 in current applications but also highlight its potential for future wide-field weak lensing surveys.

\section{Discussion and conclusion } \label{sec:discussion}
In this study, we introduce AKRA 2.0, an updated version of the original AKRA algorithm \cite{shiAKRA2024a}, designed to reconstruct accurate high-resolution $\kappa$ maps from masked shear catalogs on spherical geometries. Inspired by mapmaking methodologies from interferometry and CMB studies, we derive the the minimum-variance $\kappa$ estimator for full-sky geometry using a prior-free maximum likekihood. To optimize the computational cost, we employ the scale-splitting strategy, i.e., extracting the large-scale $\kappa$ information with the spherical-sky tool (AKRA-sphere) while obtaining the small-scale $\kappa$ information with the flat-sky tool (AKRA-flat).
Moreover, the innovative scale-splitting techniques employed in AKRA 2.0 also have potential applications in both interferometry and CMB studies.
% \hongbo{In this study, we introduce AKRA 2.0, an updated version of the original AKRA algorithm \cite{shiAKRA2024a}, designed to reconstruct high-resolution $\kappa$ maps from masked cosmic shear maps on spherical geometries. We derive the the minimum-variance $\kappa$ estimator for full-sky geometry using a a prior-free maximum likekihood, a similar methodology to that introduced in \cite{shiAKRA2024a} for flat-sky scenario. For optimizing the computational cost, we employ the scale-splitting strategy, i.e., extracting the large-scale $\kappa$ information with the spherical-sky tool (AKRA-sphere) while obtaining the small-scale $\kappa$ information with the flat-sky tool (AKRA-flat).}

AKRA 2.0 achieves high precision at both unmasked and randomly masked pixels. We evaluated its performance under various conditions, including tests on the AKRA-sphere and AKRA-flat components, and the integrated performance of the entire system. 
In unmasked pixels, the autopower spectrum ratio $C_{\ell}^{\text{rec}} / C_{\ell}^{\text{true}}$ and the cross-correlation coefficient $r_{\ell}$ consistently achieved accuracies within 1\% or better. Performance statistics such as $s$, $\rho$, and $L$ are summarized in Table \ref{tab:results}, demonstrating its high accuracy. Notably, for $L$, AKRA-sphere shows a significant reduction in contamination from masked pixels, being ${\mathcal{O}(10)}$ to ${\mathcal{O}(10^2)}$ times lower than the KS method. This capability stems from the non-local nature of the observables $\gamma_1$ and $\gamma_2$, also enabling accurate reconstruction even in small masked regions. 
% \hongbo{By testing AKRA 2.0 on simulated shear maps with various types of masks, we show it achieves high precision in recovering $\kappa$ map in the unmasked region where the autopower spectrum ratio $C_{\ell}^{\text{rec}} / C_{\ell}^{\text{true}}$ consistently achieved accuracies within 1\%, and the cross-correlation coefficient $r_{\ell}$ is nearly 1 \red{number?}. In the masked region, ...}

The current implementation of AKRA 2.0 is primarily focused on method validation with the effect of mask. It is tested under the assumption that shear maps are noise-free and that convergence maps are treated as Gaussian random fields.
However, AKRA 2.0 is equipped to automatically handle noise in shear measurements through noise covariance matrix $\bf{N}$. 
This feature will be particularly useful when the algorithm is applied to real data.
Furthermore, although AKRA 2.0 currently assumes Gaussianity for simplicity, it is fundamentally robust against the statistical nature of input data. Its non-prior characteristics allow for natural applicability to non-Gaussian distributions.

Moreover, several additional factors must be considered for practical applications in Stage IV surveys, such as varying survey depth, blending of galaxy images, spatial variations in the selection function, and shear calibration biases. We plan to incorporate realistic simulations including these factors to test and enhance the robustness of AKRA 2.0 for use in stage IV weak lensing surveys.

Future efforts will aim to implement our method into practical scenarios, addressing challenges like handling noisy shear maps and accurately treating the non-diagonal covariance matrix of the shear field.
AKRA 2.0 is well-suited for application to current surveys such as HSC, DES, KiDS and future missions such as CSST, LSST, Euclid, and Roman. Building on the accurate $\kappa$ maps produced by AKRA 2.0, high-order statistics, such as peak counts, Minkowski functionals, PDFs and field-level inferences can be robustly measured. Also, other statistics like the bispectrum, three-point correlation functions, higher moments and marked correlations can be further examined.

\acknowledgments

This work was supported by the National Key R\&D Program
of China (2023YFA1607800, 2023YFA1607801, 2020YFC2201602, 2022YFF0503403), the China Manned Space Project (\#CMS-CSST-2021-A02), China Manned Space Program (\#CMS-CSST-2025-A03),
and the Fundamental Research Funds for the Central Universities.
JY acknowledges the support from NSFC Grant No. 12203084. 
ZYS acknowledges the support from Sino-German (CSC-DAAD) Postdoc Scholarship Program, Grant No. 2023 (57678375), and Shanghai Jiao Tong University Doctoral Graduate Development Scholarship.
This work made use of the Gravity Supercomputer at the Department of Astronomy, Shanghai Jiao Tong University.
 The results in this paper have been derived using the following packages: Numpy \cite{numpy}, Scipy \cite{SciPy},
\texttt{HEALPix} \cite{healpix}, IPython \cite{ipython} and CCL \cite{pyccl}. We sincerely thank the anonymous referee for the constructive comments and suggestions.

\appendix

  \section{Notation and detail derivation in AKRA 2.0} \label{sec:app_details}
\subsection{Notation} \label{subsec:appendix_notation}
To clarify the physical quantities and symbols used in the reconstruction process, a detailed table is provided in Table \ref{tab:notation}.
The table also lists the relevant sections in which each quantity is firstly discussed.

\begin{table*}[htbp]
% \begin{ruledtabular}
\centering
\scriptsize
\caption{The physical quantities and related symbols involved in reconstruction process.}
\label{tab:notation}
\begin{tabular}{llp{7cm}}
\textbf{Quantity} & \textbf{Symbol}  &\textbf{Description}  \\
\hline
\hline
Shear components & $\gamma_{1,2}(\theta)$ (Sec.~\ref{sec:method}) & Two components $\gamma_1$ and $\gamma_2$ of the shear in real field.\\
Shear field & $\gamma(\hat{\boldsymbol{n}})$ (Sec.~\ref{sec:method}) &  Defined as $\gamma = \gamma_1 + i \gamma_2  = |\gamma| e^{2 i \phi}$ from the components $\gamma_1$ and $\gamma_2$.\\
Conjugate shear field & $\gamma^*(\hat{\boldsymbol{n}})$ (Sec.~\ref{sec:method}) & The complex conjugate of the shear field $\gamma$.\\

Spin-weighted spherical harmonic (SH) & ${ }_{s} Y_{l m}$ (Sec.~\ref{sec:method}) & \makecell[l]{The spin-weighted SH function with spin weight $s$, \\ in this work $s = \pm 2$.}\\
Coefficients of spin-2 SH & $a_{\pm 2, \ell m}$ (Sec.~\ref{sec:method}) & The coefficients of the spin-2 SH function.\\
Coefficients of E/B field SH & $a_{E/B, \ell m}$ (Sec.~\ref{sec:method}) & The coefficients of the E/B mode of the spin-0 SH function.\\
Coefficients of lensing potential SH & $\phi_{\ell m}$ (Sec.~\ref{sec:method}) & The coefficients of the lensing potential SH function.\\
Coefficients of convergence SH & $\kappa_{\ell m}$ (Sec.~\ref{sec:method}) & The coefficients of the convergence SH function.\\
Mask function & $m(\theta)$ (Sec.~\ref{sec:method}) & The mask function in real space is 1/0 for unmasked/masked pixels.\\
Masked shear field & $\gamma^{\m}(\hat{\boldsymbol{n}})$ (Sec.~\ref{sec:method}) & The masked shear field in real space.\\
Coefficients of masked shear SH & $a^{\m}_{\pm 2, \ell m}$ (Sec.~\ref{sec:method}) & The coefficients of the masked shear SH functions.\\
Mask convolution kernel & ${ }_{\pm 2}\mathbf{M}_{\ell_1 m_1 \ell m}$ (Sec.~\ref{sec:method}) & \makecell[l]{The convolution kernel from mask, with each shape of\\ $ ((\ell_{\text{max}} + 1)^2,(\ell_{\text{max}} + 1)^2) $.}\\
Coefficients of observed shear SH vector & $\boldsymbol{y}$ (Sec.~\ref{sec:method}) & The observed shear vector in SH space, with length $2(\ell_{\text{max}} + 1)^2$.\\

Reconstructed E-mode coefficients & $\boldsymbol{x}$ (Sec.~\ref{sec:method}) & \makecell[l]{The reconstructed E-mode coefficients in SH space, with length \\ $(\ell_{\text{max}} + 1)^2$.}\\

Convolution matrix & $\mathbf{A}$ (Sec.~\ref{sec:method}) & \makecell[l]{The kernel matrix transforms the convolution operation into a matrix \\multiplication operation with shape ($2(\ell_{\text{max}} + 1)^2, (\ell_{\text{max}} + 1)^2$).}\\

Noise vector & $\boldsymbol{n}$ (Sec.~\ref{sec:method}) & The noise vector, with length $2(\ell_{\text{max}} + 1)^2$. \\

Hermitian conjugate of a matrix & $\bf{A}^{\rm{T}}$ (Sec.~\ref{sec:method}) & Denotes the Hermitian conjugate of matrix A.\\

Inverse of a matrix & $\bf{A}^{-1}$ (Sec.~\ref{sec:method}) & Denotes the inverse of matrix A.\\

Regularization matrix & $\bf{R}$ (Sec.~\ref{sec:method}) & The regularization matrix, which is proportional to the identity matrix.\\
\hline
Galaxy index $i$ & in $\gamma_{1,2}^i$ (Sec.~\ref{sec:method}) & Represents an individual galaxy in our analysis.\\
Galaxy weight & $w^i$ (Sec.~\ref{sec:method})& The weight assigned to each galaxy in the analysis.\\
celestial coordinates & $(\theta_{\rm{RA}}^i, \theta_{\rm{DEC}}^i)$ (Sec.~\ref{sec:method}) & The celestial coordinates of a galaxy.\\
Galaxy density & $N_{\rm{p}}$ (Sec.~\ref{sec:method}) & The average number of galaxies per pixel.\\
Smooth kernel & $\theta_{\rm{sm}}$ (Sec.~\ref{sec:method}) & The smooth kernel applied at a scale of $\theta_{\rm{sm}}$.\\
FWHM of the smooth kernel & $R_{\rm{FWHM}}$ (Sec.~\ref{sec:method}) & The full width at half maximum of the smooth kernel.\\
Superscript $\rm{large}$ & $\gamma^{\rm{large}}$/$m^{\rm{large}}$/$\kappa^{\rm{large}}$ (Sec.~\ref{sec:method}) & The large-scale shear/mask/convergence map.\\
Grid dimensions & $N_x,N_y$ (Sec.~\ref{sec:method}) & \makecell[l]{The dimensions of each flat-sky segment, in this work we \\set $N_x = N_y = N$ for simplicity.}\\
Grid spacing & $H_x,H_y$ (Sec.~\ref{sec:method}) & The grid spacing in each segment, in this work we set $H_x = H_y = H$.\\
Grid length & $l_x,l_y$ (Sec.~\ref{sec:method}) & \makecell[l]{
The length of each segment in the x and y directions, in this work we \\set $l_x = l_y = l$.}\\
Superscript $\rm{small}$ & $\gamma^{\rm{small}}$/$m^{\rm{small}}$/$\kappa^{\rm{small}}$ (Sec.~\ref{sec:method}) & The small-scale shear/mask/convergence map.\\

\hline
Slope & $s$ (Sec.~\ref{sec:simulation}) & \makecell[l]{The slope of a linear regression model, which is used to estimate the\\ distribution of true ($\kappa^{\rm{true}}$) and reconstructed ($\kappa^{\rm{rec}}$) convergence map in \\kernel density estimate plot.}\\

Pearson correlation coefficient (PCC) & $\rho$ (Sec.~\ref{sec:simulation}) & The PCC between $\kappa^{\rm{true}}$ and $\kappa^{\rm{rec}}$.\\

The localization measure & $L$ (Sec.~\ref{sec:simulation}) &  \makecell[l]{A quantitative assessment of the degree to which the residuals are con-\\centrated within the masked regions.}\\

Power spectrum ratio & $C_{\ell}^{\rm{rec}}/C_{\ell}^{\rm{true}}$ (Sec.~\ref{sec:simulation}) & The ratio of the auto power spectrum of $\kappa^{\rm{rec}}$ to $\kappa^{\rm{true}}$.\\

Cross correlation coefficient & $r({\ell})$ (Sec.~\ref{sec:simulation}) & The cross correlation coefficient between $\kappa^{\rm{true}}$ and $\kappa^{\rm{rec}}$.\\
\hline
\end{tabular}
% \end{ruledtabular}
\end{table*}

\begin{figure*}[htbp]
    \centering
    \includegraphics[width=0.99\textwidth]{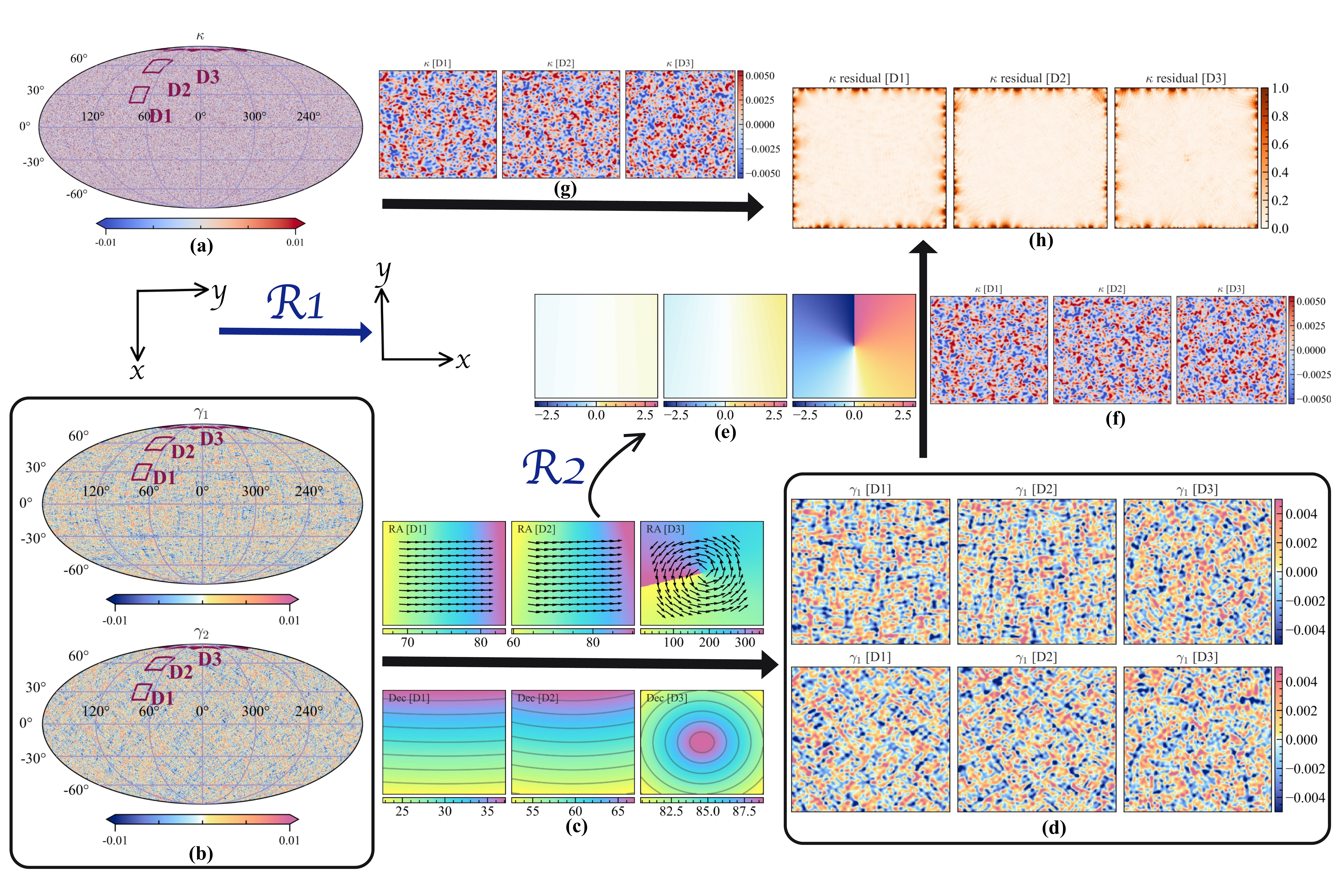}
    \caption{Validation of the transformation from spherical to flat sky representations. This comprehensive illustration details each step in the process: \textbf{(a)} The convergence map on the sphere. \textbf{(b)} Shear maps $\gamma_1$ and $\gamma_2$ on the sphere, highlighting three key regions (D1, D2, D3) at right ascension (RA) $75^\circ$ and declinations (Dec) of $30^\circ$, $60^\circ$, and $90^\circ$. \textbf{(c)} Coordinates for regions D1, D2, and D3. \textbf{(d)} Transitioned shear maps $\gamma_1$ and $\gamma_2$ to a flat sky after large-scale feature removal. \textbf{(e)} Angular corrections applied to align shear maps with the flat sky model. \textbf{(f)} Derived flat sky convergence maps using the KS method from (d). \textbf{(g)} Convergence maps for these regions directly extracted from the spherical map. \textbf{(h)} Comparison of convergence maps highlighting discrepancies between the KS method results and direct spherical extractions. This process requires two angular corrections, represented in $\bm{\mathcal{R}}_1$ and $\bm{\mathcal{R}}_2$, which will be detailed in Appendix \ref{subsec:transform}.}
    \label{fig:sphere_to_flat}
\end{figure*}

\subsection{Calculate the Mask Convolution Kernel} \label{subsec:appendix_convoltion}
In spherical harmonic (SH) analysis, the calculation of the product of two SH functions within the SH basis involves tripling coefficients.
To derive the $i$-th coefficient of the SH projection of the product $c(\hat{\boldsymbol{n}}) = m(\hat{\boldsymbol{n}}) b(\hat{\boldsymbol{n}})$, we express it mathematically as: 
\begin{equation}
    \begin{aligned} 
        c_{i} 
        & =\int_{\Omega_{4 \pi}} y_{i}(\hat{\boldsymbol{n}}) m(\hat{\boldsymbol{n}}) b(\hat{\boldsymbol{n}}) d \hat{\boldsymbol{n}} \\
        & =  m_{j} b_{k} \int_{\Omega_{4 \pi}} y_{i}(\hat{\boldsymbol{n}}) y_{j}(\hat{\boldsymbol{n}}) y_{k}(\hat{\boldsymbol{n}}) d \hat{\boldsymbol{n}} \\ & = m_{j} b_{k} C_{i j k}
    \end{aligned}
    \label{eq:double_product_3j}
\end{equation}
where $C_{ijk}$ denotes the tripling coefficients, and $y_i(\hat{\boldsymbol{n}}), y_j(\hat{\boldsymbol{n}}), y_k(\hat{\boldsymbol{n}})$ are the SH basis. The indices $i, j, k$ represent the modes of the SH transform, while $m_j$ and $b_k$ are the SH coefficients of $m(\hat{\boldsymbol{n}})$ and $b(\hat{\boldsymbol{n}})$, respectively. 

If $m(\hat{\boldsymbol{n}})$ is known, the computation can be simplified using a transfer matrix $\mathbf{M}$. This approach reduces the complexity of calculating the projection of the product directly and is described by: 
\begin{equation}
    \begin{aligned} 
        c_{i} 
        % & =\int_{\Omega_{4 \pi}} y_{i}(\hat{\boldsymbol{n}}) m(\hat{\boldsymbol{n}}) b(\hat{\boldsymbol{n}}) d \hat{\boldsymbol{n}} \\
        %  & =\int_{\Omega_{4 \pi}} y_{i}(\hat{\boldsymbol{n}})\left( b_{k} y_{k}(\hat{\boldsymbol{n}})\right) m(\hat{\boldsymbol{n}}) d \hat{\boldsymbol{n}} \\ 
        % & = b_{k} \int_{\Omega_{4 \pi}} y_{i}(\hat{\boldsymbol{n}}) y_{k}(\hat{\boldsymbol{n}}) m(\hat{\boldsymbol{n}}) d \hat{\boldsymbol{n}} \\
         & = \mathbf{M}_{i k}b_{k}.
    \end{aligned}
    \label{eq:double_product_transfer}
\end{equation}
This matrix $\mathbf{M}_{ik}$ encapsulates the impact of the known function $m(\hat{\boldsymbol{n}})$ on the resulting coefficients by relating the SH indices $i$ and $k$ through the mask function $m(\hat{\boldsymbol{n}})$.

When applying a mask to a spin-weighted field, the mask must also adopt the spin characteristics of the field, necessitating the use of a spin-weighted SH transform. The modification for these spin-weighted fields is expressed as:
\begin{equation}
    \mathbf{M}_{ik} = \int_{\Omega_{4\pi}} {}_{s}Y_{i}(\hat{\boldsymbol{n}}) \, {}_{s}Y_{k}(\hat{\boldsymbol{n}}) \, m(\hat{\boldsymbol{n}}) \, d\hat{\boldsymbol{n}},
\end{equation}
where ${}_{s}Y_i(\hat{\boldsymbol{n}})$ and ${}_{s}Y_k(\hat{\boldsymbol{n}})$ are the spin-weighted SH basis functions, and $s$ is the spin weight. 
% The spin number $s$, such as $\pm 2$ for shear fields, dictates the type of field deformation influenced by interactions and is crucial for preserving the correct transformation properties under rotations.

The necessity of using transfer kernel in the convolution of spin-weighted shear fields highlights the non-local nature of shear data. How to get the transfer kernel is a key issue in the AKRA 2.0 algorithm. The transfer kernel can be obtained by implementing a delta function approach. This method allows for the precise calculation of the mask convolution kernel, ensuring that the mask's effects are accurately captured and corrected in the final mass map reconstructions.

The \texttt{healpy} library plays a crucial role in SH analysis for cosmological data. The \texttt{hp.map2alm} function computes SH coefficients $a_{lm}$ from a sky map, but only for $m \geq 0$. This limitation is due to the real nature of the field, requiring symmetry to obtain coefficients for negative $m$:
$$a_{l -m} = (-1)^m a_{lm}^*, $$
where $a_{lm}^*$ is the complex conjugate of $a_{lm}$.
The inverse operation, performed by \texttt{hp.alm2map}, transforms these coefficients back into a spatial map.

To understand how masks influence SH fields, we use a delta function approach, outlined as follows:

\begin{enumerate}
    \item {Delta Function Initialization}: Start by defining a delta function at a specific SH space point by setting a chosen coefficient $a_{lm} = 1$ and all others to zero. This represents a point source in harmonic space. Adjust for both positive and negative $m$ due to the \texttt{healpy} limitation:
    \begin{itemize}
        \item Set $a_{lm} = 1$ and $a_{l -m} = (-1)^m$, enabling the calculation of $M_1 = M_{lm} + (-1)^m M_{l -m}$.
        \item For imaginary parts, set $a_{lm} = i$ and $a_{l -m} = -(-1)^m i$, resulting in $M_2 = i(M_{lm} - (-1)^m M_{l -m})$.
    \end{itemize}

    \item {Transformation and Mask Application}: Use \texttt{hp.alm2map} to convert these coefficients back into a spatial map, then multiply by the observational mask to simulate masking effects.

    \item {Masked Map Transformation}: Convert the masked map back into SH coefficients with \texttt{hp.map2alm}, analyzing how the mask alters data for specific $ (l, m) $ modes.

    \item {Kernel Extraction}: Repeat across various $ (l, m) $ modes to compile data on the mask's impact, forming a convolution kernel dataset essential for correcting mask effects in AKRA 2.0 analysis.
\end{enumerate}

These setups help in effectively extracting both real and imaginary components:
$$M_{lm} = \frac{(M_1 + M_2/i)}{2},$$
$$(-1)^m M_{l-m} = \frac{(M_1 - M_2/i)}{2}.$$

This methodology enables the reconstruction of the mask convolution kernel for all $m$ values, providing a robust framework for accurate cosmological data analysis.

\subsection{Transformation from spherical to flat-sky coordinates} \label{subsec:transform}
In AKRA 2.0, we integrate the AKRA-sphere and AKRA-flat algorithms to efficiently handle both large-scale and small-scale components of the data. In this section, we will detail the validation process from spherical to flat-sky coordinates, which is foundation to perform the scale-splitting method.

Visual results, illustrated in Fig.~\ref{fig:sphere_to_flat}, offer a clear depiction of each step in the transformation process. This vlaidation starts from the initial spherical map to the part of the shear maps, and then we obtain the flat-sky convergence maps under the assumption of periodic boundary conditions. The last panel (h) illustrates the discrepancies that arise when employing the KS method under these boundary conditions compared to direct spherical extractions.
This analysis reveals that maximum residual errors are predominantly located at the margins of the reconstructed convergence map, while the central regions maintain high accuracy.  
To effectively address these boundary-induced discrepancies, AKRA 2.0 specifically targets and refines the treatment of edge areas in the mask, reducing boundary values to zero. 
This adjustment ensures that the central regions of the maps, which contain the most reliable data, are prioritized in the analysis, enhancing the overall accuracy of the convergence map reconstruction.

The transformation employs a mathematical framework involving rotation matrices to adjust the shear components accurately, which are spin-weighted fields. The rotation matrix applied is given by:
\begin{equation}
\begin{pmatrix}
        \tilde{\gamma}_{1} \\
        \tilde{\gamma}_{2}
    \end{pmatrix}
    = \bm{\mathcal{R}}(\phi_{0})
    =
    \begin{pmatrix}
        \cos 2 \phi_{0} & \sin 2 \phi_{0} \\
        -\sin 2 \phi_{0} & \cos 2 \phi_{0}
    \end{pmatrix}
    \begin{pmatrix}
        \gamma_{1} \\
        \gamma_{2}
    \end{pmatrix},
\end{equation}
where $\phi_0$ is the rotation angle required to align the shear maps with the flat sky coordinates. In this equation, $\tilde{\gamma}_{1}$ and $\tilde{\gamma}_{2}$ represent the transformed shear components in the rotated coordinate system. 

This process requires two following angular corrections.
Firstly, the transition from spherical coordinates $(\bm{e}_{\theta}, \bm{e}_{\phi})$ of shear maps to the flat sky coordinates $(\bm{x}_{\text{RA}}, \bm{y}_{\text{Dec}})$ necessitates a rotational adjustment of $90^\circ$. So the first rotation correction is :
$$ 
\bm{\mathcal{R}}_1 = \bm{\mathcal{R}}(90^\circ) =
    \begin{pmatrix}
        -1 & 0 \\
        0 & -1
    \end{pmatrix}.
$$
As declinations increase, especially beyond $30^\circ$, the spherical curvature becomes more pronounced, complicating the direct flat-sky projection. Then we generate the grid dimensions centered at the equator, and rotate regions near the equator to align with their corresponding declinations. Moreover, as the right ascension axis converges towards the poles, further rotational adjustments are required. These are calculated based on the angular gradient of the right ascension axis at each cell, illustrated in Fig.~\ref{fig:sphere_to_flat} (c). The corresponding rotation matrix is:
$$
\bm{\mathcal{R}}_2 = \bm{\mathcal{R}}(\phi_{\rm RA}) =
    \begin{pmatrix}
        \cos 2 \phi_{\rm RA} & \sin 2 \phi_{\rm RA} \\
        -\sin 2 \phi_{\rm RA} & \cos 2 \phi_{\rm RA}
    \end{pmatrix},
$$
where $\phi_{\rm RA}$ is the angle necessary to correct for the RA variations due to declination.  
Fig.~\ref{fig:sphere_to_flat} illustrating regions D1, D2, and D3 demonstrate the transformation's impact. Each region, affected by its specific declination, showcases the necessity and effectiveness of the tailored rotational adjustments. 
This correction is particularly critical near the poles, as evidenced by the adjustments needed in regions like D3 (shown in Fig.~\ref{fig:sphere_to_flat}).

\section{Testing AKRA-sphere using random masks} \label{sec:app_simulation_sphere}
In this section, we compare the performance of AKRA 2.0 at large scale only (AKRA-sphere) and KS algorithms using observed shear fields with known $\kappa$ fields. We use the same simulation setup as in Sec.~\ref{subsec:sphere_performance}, with $N_{\rm{side}} = 128$ and $\ell_{\rm{max}} = 195$. 
We will focus on the impact of the random mask on the reconstruction results. Specifically, we simulate using the following masks: 
\begin{itemize}
    \item Random masks with varying rates of masked pixels: 10\%, 30\%, and 50\%.
    \item Random masks with varying sizes of masked pixels with fixed 40\% masked pixels.
\end{itemize}

\begin{figure*}[htbp]
    \centering
    \includegraphics[width=0.85\textwidth]{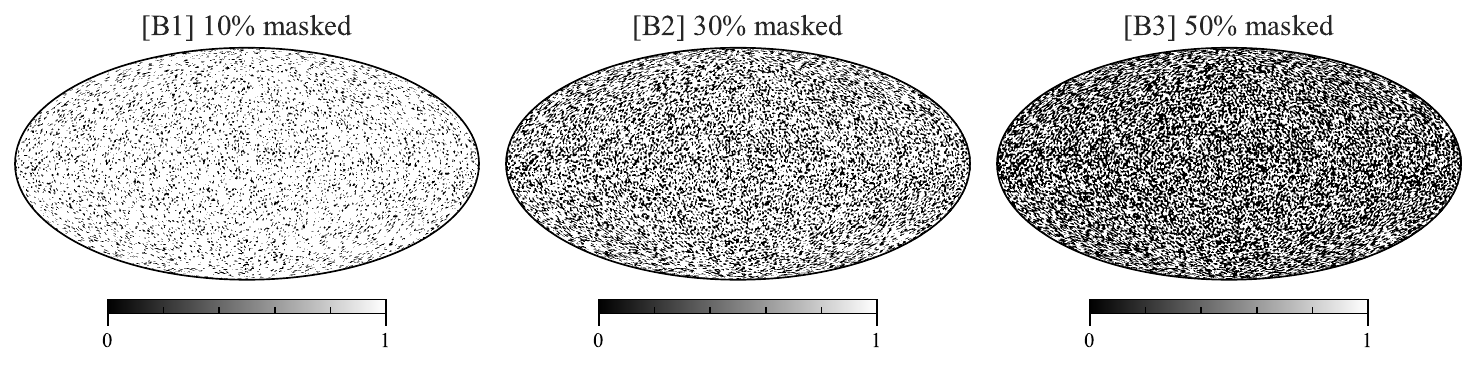}
    \caption{The mask is generated by randomly selecting pixels from $m(\hat{\boldsymbol{n}})$ and setting them to zero. In panels B1 to B3, we present masks with varying rates of masked pixels: 10\%, 30\%, and 50\%, respectively. }
    \label{fig:random_g1g2Mask}
\end{figure*}

\begin{figure*}[htbp]
\centering
    \subfigure[]{\includegraphics[width=0.85\textwidth]{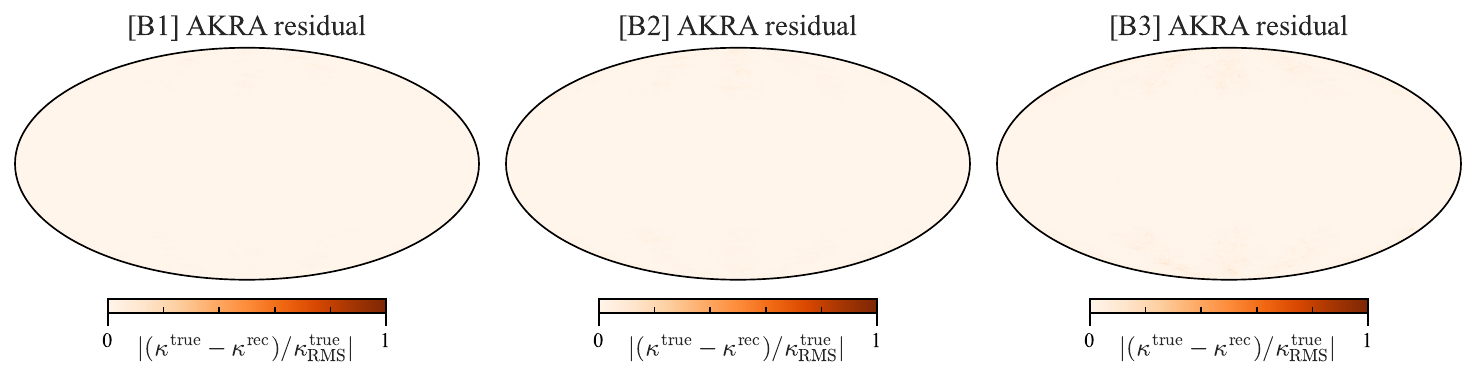}} \\
    \subfigure[]{\includegraphics[width=0.85\textwidth]{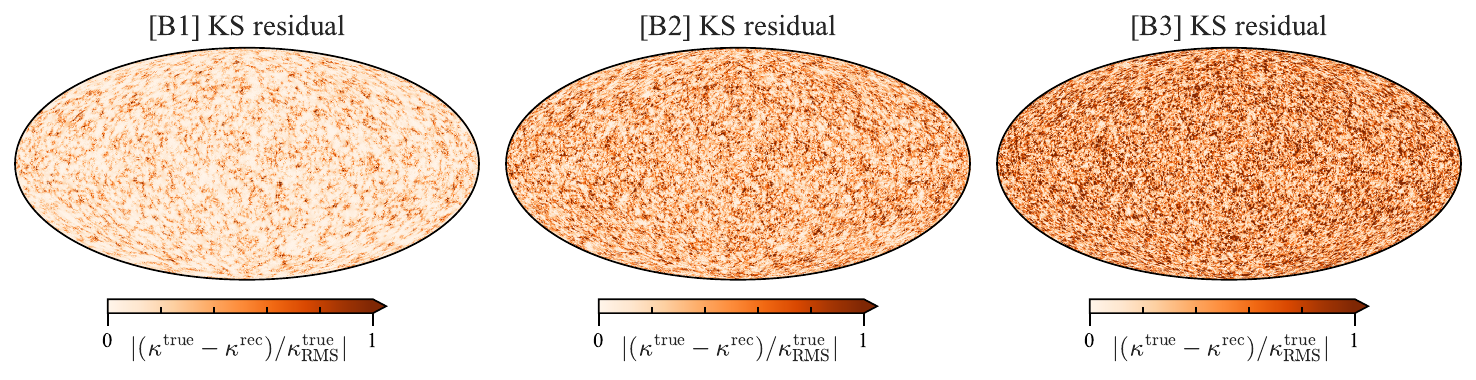}}
    \caption{The residual maps, normalized by the r.m.s. of the true signal, are presented for AKRA 2.0 (a) and the KS method (b). From left to right, the three columns correspond to the three masks with varying rates of masked pixels shown in Fig.~\ref{fig:random_g1g2Mask}. The KS method exhibits significant residuals across all cases (B1 to B3) due to the presence of masked pixels. In contrast, AKRA 2.0 demonstrates consistent robustness.}
    \label{fig:random_g1g2Mask-results} 
\end{figure*}

\begin{figure*}[htbp]
    \centering
    \includegraphics[width=0.85\textwidth]{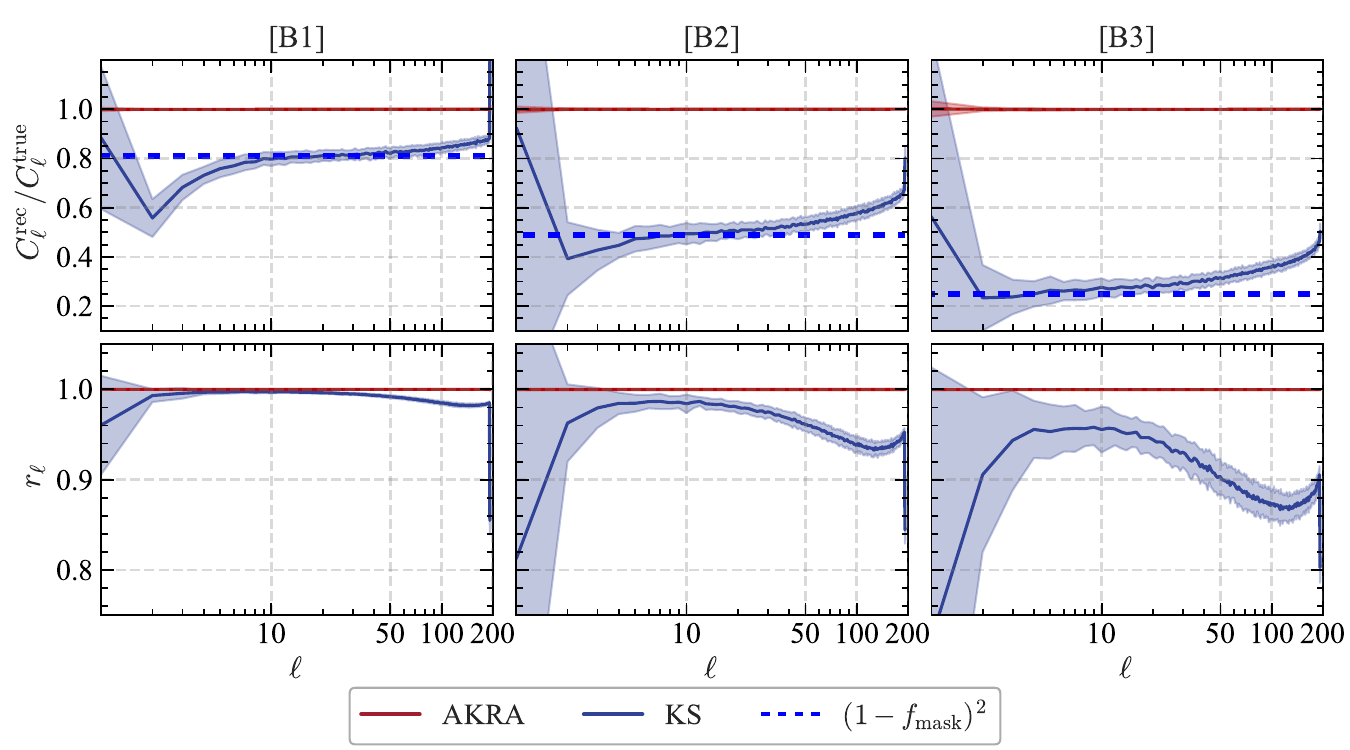}
    \caption{Results for 100 realizations using random masks. Top: power spectrum ratio for 100 realizations (only use {\bf unmasked pixels} and set $\kappa$ in masked pixels to be zero). The blue and red regions represent the 1$\sigma$ confidence interval.   
    Bottom: cross-correlation coefficient between the reconstructed and the true convergence map.  Notably, we did not correct the bias in the power spectrum derived from the KS method due to varying mask rates, as the effects on different scales are not uniform. In contrast, AKRA 2.0 recovers the true power spectrum in the simulation with high accuracy, yielding the power spectrum ratio and cross-correlation coefficient that deviate from  $1$ by less than $1\%$, respectively.}
    \label{fig:random_rk}
\end{figure*}

\begin{figure*}[htbp]
    \centering
    \includegraphics[width=0.85\textwidth]{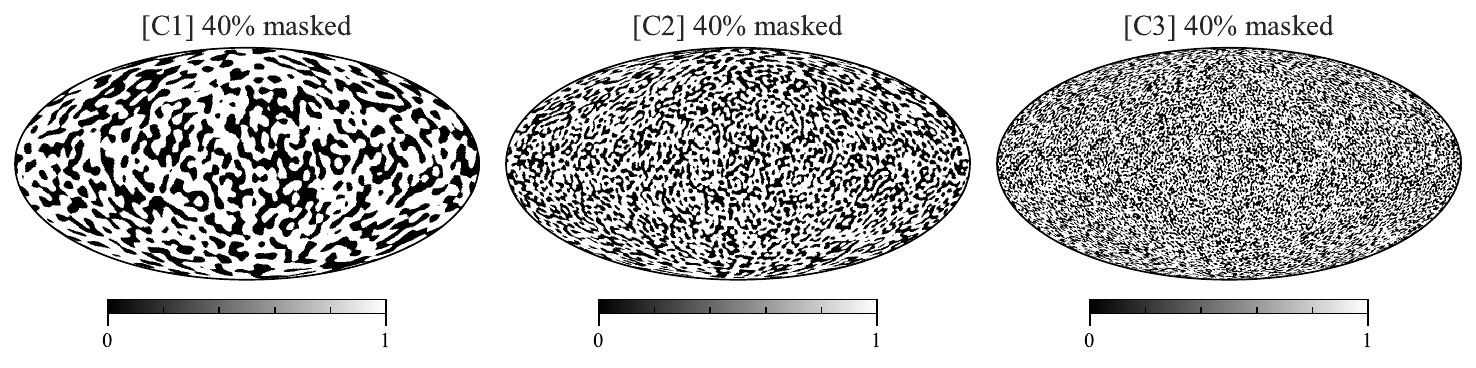}
    \caption{Similar to Fig.~\ref{fig:random_g1g2Mask}, but with a different scale for the mask and a fixed mask rate of $40\%$. The mask maps are labeled C1 to C3, each generated from a flat power spectrum with $l_{\text{max}} $ values of 50, 100, and 192 ($1.5 \times N_{\text{side}}$), respectively.}
    \label{fig:random_g1g2Mask2}
\end{figure*}

\begin{figure*}[htbp]
\centering
    \subfigure[]{\includegraphics[width=0.85\textwidth]{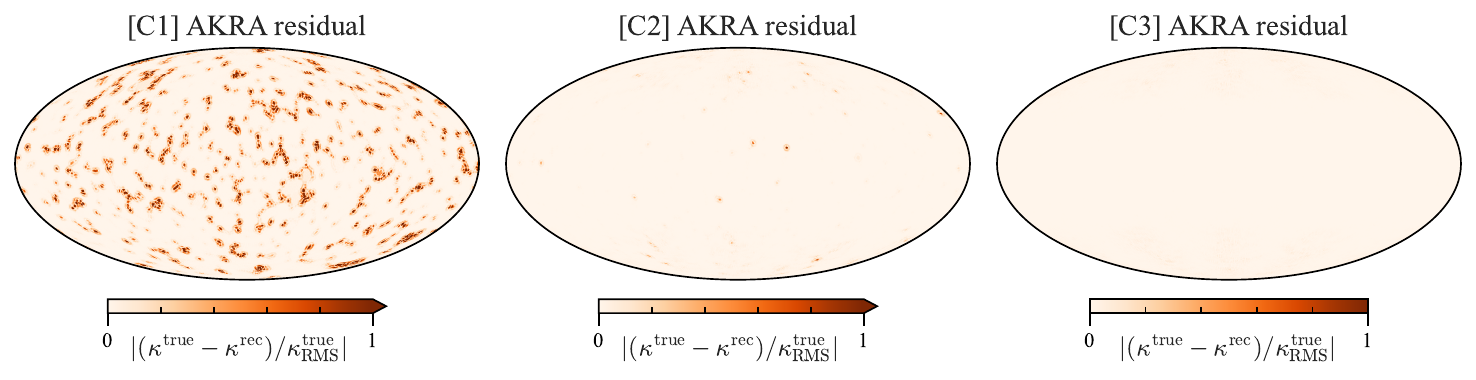}} \\
    \subfigure[]{\includegraphics[width=0.85\textwidth]{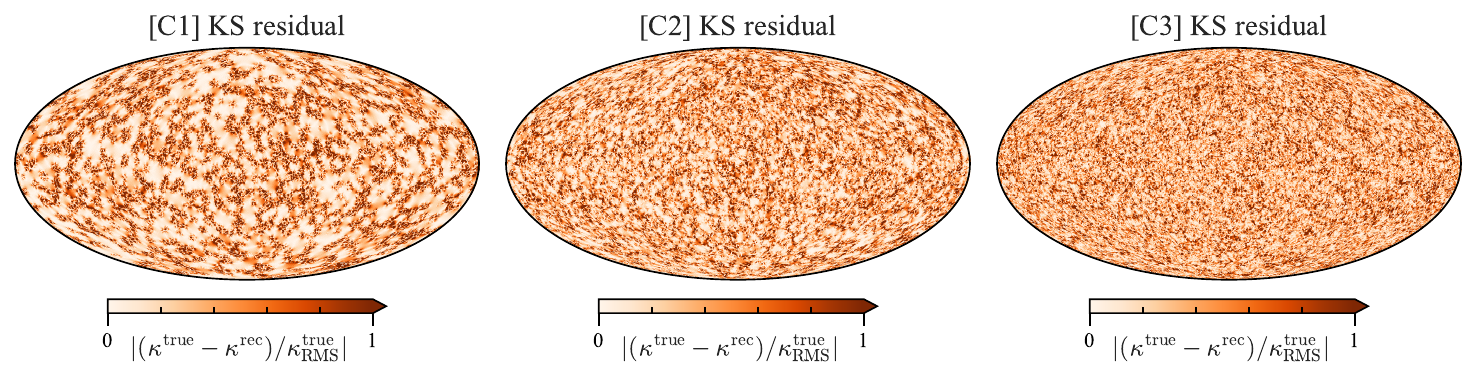}}
    \caption{Similar to Fig.~\ref{fig:random_g1g2Mask-results}, but with diffenet mask presented in Fig.~\ref{fig:random_g1g2Mask2}. The normalized residuals produced by AKRA 2.0 are markedly smaller than those from the KS method. In case C1, the normalized residuals approach zero in the unmasked regions, indicating a high level of accuracy. For cases C2 and C3, where the masked pixels are less clustered, the residuals are zero in both masked and unmasked areas, demonstrating the robustness of the AKRA 2.0 method in handling small scale mask scenarios.}
    \label{fig:random_g1g2Mask-results2} 
\end{figure*}

\begin{figure*}[htbp]
    \centering
    \includegraphics[width=0.85\textwidth]{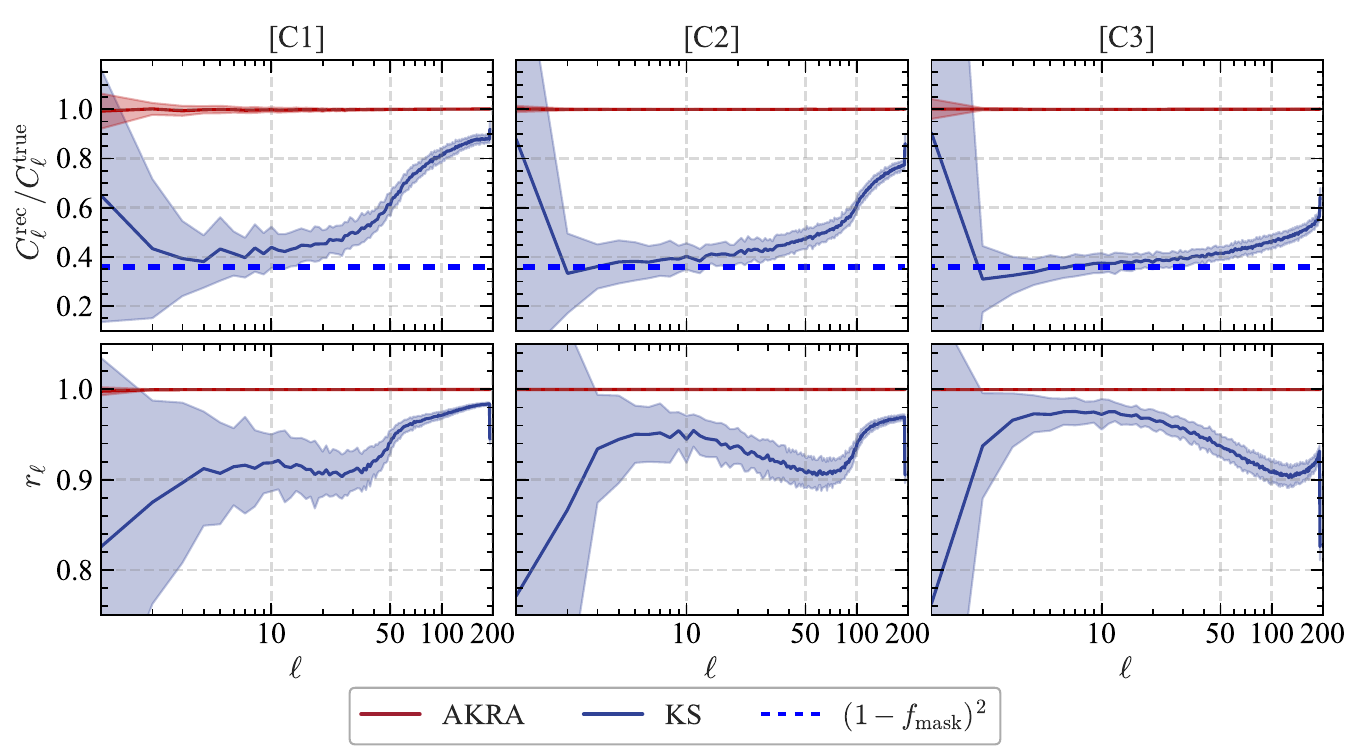}
    \caption{Similar to Fig.~\ref{fig:random_rk}, but with diffenet mask scales with a fixed mask fraction of $40\%$ presented in Fig.~\ref{fig:random_g1g2Mask2}. The power spectrum ratio and cross-correlation coefficient from AKRA 2.0 outperform those from the KS method. The blue dashed line represents the $(1-f_{\text{mask}})^2$ scaling factor. 
    Notably, in the KS method, a constant scaling factor cannot effectively correct the power spectrum due to the non-uniformity of the mask effect across different scales. In contrast, AKRA 2.0 enables us to calculate the power spectrum with high accuracy, eliminating the need for any corrections.}
    \label{fig:random_rk2}
\end{figure*}

\subsection{Varying fractions of masked pixels}
% The random mask is generated from a flat power spectrum using a Gaussian random field, with a threshold applied to set values below this threshold to 0. 
% The three panels in Fig. \ref{fig:random_g1g2Mask} depict masks with varying masked pixel rates of 10\% (B1), 30\% (B2), and 50\% (B3).

We present the residual maps normalized by r.m.s. $| \kappa^{\text{true}} - \kappa^{\text{rec}}/\kappa^{\text{true}}_{\text{RMS}}|$ obtained from AKRA 2.0 (panel a) and the KS method (panel b) in Fig.~\ref{fig:random_g1g2Mask-results}. The KS method yields substantial residuals due to masked pixels across all cases (B1-B3). In contrast, AKRA 2.0 showcases remarkable robustness, consistently producing minimal residuals in both masked and unmasked regions.

In Fig.~\ref{fig:random_rk}, a clear distinction arises between the performance of the AKRA 2.0 and KS methods. The power spectrum of the reconstructed $\kappa$ map from the KS method is highly sensitive to the fraction of masked pixels. Conversely, the power spectrum ratio from AKRA 2.0 remains close to 1 for mask fractions up to 50\%, highlighting its resilience to variations in mask effect. 
% Notably, even in case B3, where half of the pixels are unmasked, both the power spectrum ratio and the cross-correlation coefficient for AKRA 2.0 are close to 1.
% These suggest that AKRA 2.0 outperforms the KS method in reconstructing the $\kappa$ map from the shear map, particularly in small masked pixels.

\subsection{Varying sizes of masked pixels}

% In this section, we test our algrithm utilizing masks of varying scales, each designed to maintain a fixed pixel coverage of 40\%. Fig.~\ref{fig:random_g1g2Mask2} illustrates the masks generated from a flat power spectrum, characterized by $ l_{\text{max}} $ values of 50, 100, and 192 ($ 1.5 \times N_{\text{side}} $). These masks are denoted as C1, C2, and C3, respectively, reflecting the increasing complexity associated with larger scales.

We present a comparison of the residual maps produced by the KS and AKRA 2.0 algorithms across the various mask scales in Fig.~\ref{fig:random_g1g2Mask-results2}. The KS method reveals significant residuals in all cases (C1 to C3), primarily due to the disruptive influence of masked pixels, which complicates the reconstruction process. In contrast, AKRA 2.0 demonstrates exceptional robustness, yielding minimal residuals in unmasked regions. Notably, for smaller mask scale (cases C2 and C3), the residuals are effectively reduced to zero within the masked areas. In the case of the larger scale (C1), although the residuals remain relatively small, they are still significantly lower compared to those generated by the KS method. This performance underscores AKRA 2.0's ability to excel at the boundaries of circular masks, successfully recovering several pixels adjacent to these regions.

Fig.~\ref{fig:random_rk2} provides additional insights by displaying the power spectrum ratio and cross-correlation coefficients for the masks under 100 realizations, focusing specifically on unmasked pixels. Under the fixed masked fraction of 40\%, AKRA 2.0 consistently surpasses the performance of KS. Notably, the reconstructed $ C_\ell^{\rm{rec}} $ values from AKRA 2.0 achieve an accuracy of better than 1\%, while the cross-correlation coefficients $ r_\ell $ remain close to 1 across all examined cases.

\begin{figure*}
    \centering
    \includegraphics[width=0.95\textwidth]{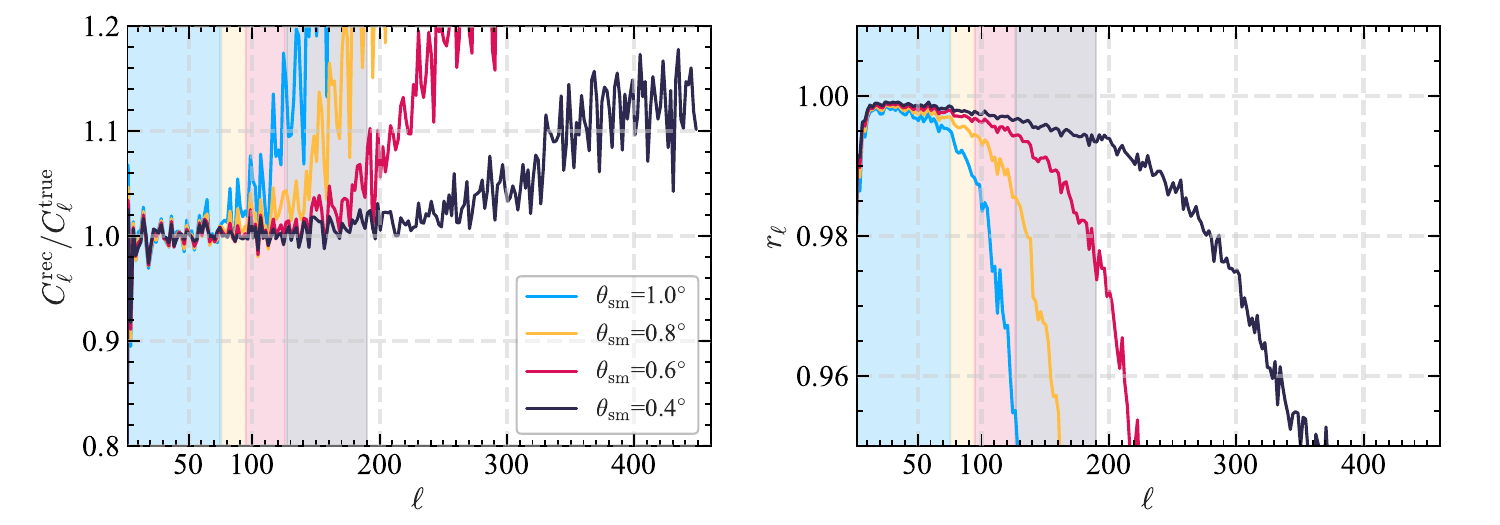}
    \caption{Reconstructed $\kappa$ map results of the CSST-like mask using AKRA 2.0 at large scales are presented. We applied different smoothing scales to the shear catalog to generate the large-scale shear field. The true $\kappa$ map is also smoothed with the corresponding kernels. 
     The smoothing scales corresponding to the blue, yellow, red, and black lines are $1.0^\circ$, $0.8^\circ$, $0.6^\circ$, and $0.4^\circ$ respectively. The $\ell_{\text{max}}$ is set to 180, 225, 300, and 450 for each line, respectively. The shaded areas indicate the range where $\ell < \ell_{\text{FWHM}}$ for each smoothing scale, highlighting where the cross-correlation coefficient's deviation from 1 is less than 1\%.}
    \label{fig:CSST_lowscale}
\end{figure*}

\section{Calculation discussion} \label{sec:app_calculation}
As discussed in Sec.~\ref{sec:method}, the matrix operations involved in Eq. \ref{eq:kappa_estimator2} present significant computational challenges. This section offers essential insights for readers keen on undertaking these calculations.

We employ the \texttt{HEALPix} scheme \citep{healpix} to pixelize the sky, where the resolution is dictated by the variable $N_{\rm{side}}$, leading to a total of $N_{\rm{pix}} = 12 \times N_{\rm{side}}^2$ pixels. The maximum multiple moment, $\ell_{\rm{max}}$, is set to $3 \times N_{\rm{side}} - 1$, but to mitigate aliasing effects, we limit our reconstruction to $\ell_{\rm{max}} \leq 1.5  N_{\rm{side}}$. This restriction ensures the precision of the $a_{\ell m}$ and $C_{\ell}$ calculations within \texttt{HEALPix}, and when using the \texttt{healpy} routines \texttt{synfast} and \texttt{alm2map} for generating full-sky shear and convergence maps. 

The operation of the matrix $\mathbf{A}$, which can become particularly large as the resolution increases, presents substantial computational hurdles. For instance, at a resolution of $1^{\circ}$, $\ell_{\rm{max}}$\footnote{The $\ell_{\rm{max}}$ is also determined by the Nyquist frequency of the smoothing kernel applied, not just by $N_{\rm{side}}$. When a $1^{\circ}$ smoothing kernel is used, it limits $\ell_{\rm{max}}$ to capture only the modes smaller than $\pi/\theta_{\rm{sm}}$ effectively represented by the kernel.}
approaches 180, which results in $\mathbf{A}$ having dimensions approximately $(6.4 \times 10^4, 3.2 \times 10^4)$. Effective memory management is crucial during this matrix construction to prevent computational overload.

The direct inversion of $\left(\mathbf{A}^{\rm{T}} \mathbf{N}^{-1} \mathbf{A}\right)$ represents a significant challenge, especially under ill-posed conditions such as those introduced by large-scale masks. A viable approach to better understand and manage this inversion is through eigendecomposition:
$$\left(\mathbf{A}^{\rm{T}} \mathbf{N}^{-1} \mathbf{A}\right)^{-1} = \mathbf{V} \boldsymbol{\Lambda} \mathbf{V}^{\rm{T}},$$ 
where $\mathbf{V}$ is composed of the eigenvectors and $\boldsymbol{\Lambda}$, a diagonal matrix, includes the eigenvalues. This analysis is critical as it elucidates how the deconvolution matrix amplifies or attenuates different components of the signal. In ill-posed scenarios, some eigenvalues of $\left(\mathbf{A}^{\rm{T}} \mathbf{N}^{-1} \mathbf{A}\right)$ are typically small, which can lead to numerical instability during inversion. A more detailed exploration of eigendecomposition and its implications for the AKRA method has been discussed in the context of AKRA-flat \citep{shiAKRA2024a}.

To ensure numerical stability, we incorporate a regularization matrix, $\mathbf{R} = \epsilon^2 \mathbf{I}$, into Eq. \ref{eq:kappa_estimator2}. This regularization helps by adding a small value $\epsilon^2$ to the eigenvalues, enhancing the robustness of the inversion process. 
The regularization matrix $\mathbf{R}$ is a diagonal matrix of the same dimensions as $\mathbf{A}^{\rm{T}} \mathbf{N}^{-1} \mathbf{A}$, serving several crucial roles. Firstly, it stabilizes the inversion process, ensuring the reliability of the computed inverse matrix. Secondly, it can suppress noise in the reconstructed maps, enhancing the quality of the results \citep{Zheng.etal2016}. Thirdly, $\mathbf{R}$ facilitates the incorporation of prior knowledge about the convergence field, allowing for the incorporation of prior knowledge regarding the convergence field \citep{Zheng.etal2016}.

In our implementation, we have chosen $\epsilon^2 = 10^{-4}$ to keep numerical stability while adhering to a non-prior approach regarding the convergence field. 
The essence of the AKRA 2.0 algorithm lies in its use of the convolution kernel $\mathbf{A}$ to manage the non-local relationship between the masked shear field and the true convergence field. This approach allows for the direct reconstruction of the convergence field from the shear measurements, effectively handling the complexities introduced by the masked observational data. AKRA 2.0's ability to operate without imposing any priors sets it apart from traditional mass-mapping methods and underscores its suitability for precise and unbiased reconstructions of the convergence field.

Inverting the matrix $\mathbf{A}^{\rm{T}} \mathbf{N}^{-1} \mathbf{A}$, particularly when it exceeds dimensions of ($10^4, 10^4$), presents formidable computational challenges. Direct inversion of such large-scale matrices often becomes impractical due to computational constraints. To address these challenges, this work employs iterative methods to simplify the inversion process and introduces a novel split-scale method to effectively manage the matrix size.

Historical efforts in large matrix inversion demonstrate that iterative methods offer a practical alternative, bypassing the need for direct matrix inversion. Notable iterative techniques include the Newton-Raphson and conjugate gradient methods, which have been effectively utilized in CMB analysis and power spectrum estimation \citep{Bond1998, Seljak1998, Hu2001, Oh1999}. Additionally, innovative approaches such as the iterative scheme by Ref.~\citep{Pen2003} and renormalization group methods outlined by Refs.\citep{McDonald2019a, McDonald2019b} further expand the toolkit for handling complex matrix operations.

In this work, we will utilize the conjugate gradient method to solve the inverse of the matrix $\mathbf{A}^{\rm{T}} \mathbf{N}^{-1} \mathbf{A} + \mathbf{R}$. The regularization matrix $\mathbf{R}$ acts to stabilize the solution by conditioning the matrix, thereby ensuring that the conjugate gradient method converges more rapidly and reliably \citep{Maraio2023}. 
In Appendix \ref{subsec:appendix_higher_resolution}, we also explore the capabilities of this method by tackling the inversion of a matrix of $(2 \times 10^5, 2 \times 10^5) $ size (corresponding to $\ell_{\rm{max}} = 450$), which is close to the practical limit of current computational resources. Our findings indicate that the conjugate gradient method is highly effective up to certain scales, but handling larger matrices remains challenging due to hardware limitations.

\subsection{Extending AKRA 2.0 to higher resolution} \label{subsec:appendix_higher_resolution}

In this section, we discuss the application of AKRA 2.0 for reconstructing $\kappa$ maps at large scales, as detailed in Sec.~\ref{sec:simulation}. As shown in Fig.~\ref{fig:CSST_lowscale}, we applied smoothing scales of $1.0^\circ$, $0.8^\circ$, $0.6^\circ$, and $0.4^\circ$ to the shear catalog to generate the large-scale shear field. The corresponding maximum multiple moments ($\ell_{\text{max}}$) were set at 180, 225, 300, and 450. 
% These settings ensured that deviations from 1 in the cross-correlation coefficient, observed for $\ell < \ell_{\text{FWHM}}$, are predominantly influenced by large-scale contributions.
As $\ell_{\rm{max}}$ reaches 450, the computational demands increase substantially, with the inverse matrix required growing to approximately $(2 \times 10^5, 2 \times 10^5)$ elements. To manage this significant computational cost, we employ the conjugate gradient method, which accelerates the calculations by iteratively converging to the solution. 
% This method significantly reduces the number of operations required compared to traditional direct matrix inversion methods. Despite these advancements, such large matrix sizes are near the upper limit of our current computational capabilities, delineating a boundary for feasible calculations.

In Fig.~\ref{fig:CSST_results}, a smooth kernel of $1.0^\circ$ was used to process the large-scale shear catalog, with a grid resolution of $0.05^\circ$ for the small-scale shear catalog, achieving high-precision results up to $\ell_{\text{max}} \sim 1500$. For achieving higher resolutions, smaller smoothing scales at the large-scale analysis are employed, as depicted in Fig.~\ref{fig:CSST_lowscale}. For instance, with an $\ell_{\rm{max}}$ of 450 and a grid resolution of $0.02^\circ$, it is feasible to attain high-precision results at $\ell_{\rm{max}} \sim 3700$.

While this work has primarily implemented a dual-scale strategy, incorporating three or more hierarchical levels of scaling is theoretically feasible. By refining these methods, we anticipate achieving unprecedented precision in cosmological mass mapping, further unlocking the potential of cosmic shear surveys to probe the underlying matter distribution.

\bibliographystyle{JHEP}
\bibliography{biblio.bib}

\end{document}